# From Data Creator to Data Reuser: Distance Matters

Christine L. Borgman, University of California, Los Angeles. Christine.Borgman@UCLA.edu, Corresponding author. ORCID: 0000-0002-9344-1029
Paul T. Groth, University of Amsterdam, p.t.groth@uva.nl, ORCID: 0000-0003-0183-6910

**ABSTRACT**

Sharing research data is necessary, but not sufficient, for data reuse. Open science policies focus more heavily on data sharing than on reuse, yet both are complex, labor-intensive, expensive, and require infrastructure investments by multiple stakeholders. The value of data reuse lies in relationships between creators and reusers. By addressing knowledge exchange, rather than mere transactions between stakeholders, investments in data management and knowledge infrastructures can be made more wisely. Drawing upon empirical studies of data sharing and reuse, we develop the metaphor of *distance* between data creator and data reuser, identifying six dimensions of distance that influence the ability to transfer knowledge effectively: domain, methods, collaboration, curation, purposes, and time and temporality. We explore how social and socio-technical aspects of these dimensions may decrease – or increase – distances to be traversed between creators and reusers. Our theoretical framing of the distance between data creators and prospective reusers leads to recommendations to four categories of stakeholders on how to make data sharing and reuse more effective: data creators, data reusers, data archivists, and funding agencies. 'It takes a village' to share research data – and a village to reuse data. Our aim is to provoke new research questions, new research, and new investments in effective and efficient circulation of research data; and to identify criteria for investments at each stage of data and research life cycles.

## 1 Why, when, how, and for whom should data be shared?

Sharing research data is now standard practice in most academic disciplines, variously required or recommended by governments, funding agencies, and scholarly journals (Barker et al., 2022; Directorate-General for Research and Innovation, European Commission et al., 2021, 2021; National Health and Medical Research Council, 2018; National Institutes of Health, 2017; Office of The Director, National Institutes of Health, 2020; Wilkinson et al., 2016). Data sharing takes many forms, whether by depositing datasets in public archives, attaching datasets to research publications, or making them available on demand. Public policy arguments for data sharing are predicated on principles of transparency or reuse (National Academies of Sciences, 2019, 2021; O'Grady, 2023; Weeden, 2023). Transparency is the principle that scholars should show their work for inspection by others to verify, validate, or reproduce findings. Reuse is based on the notion that datasets are building blocks of research that can be redeployed in future investigations.

Sharing research data is necessary, but not sufficient, for data reuse. Knowledge exchange requires much more than a transactional relationship in which creators deposit and reusers retrieve and redeploy. We argue that the *distance* to be traversed between data creators and reusers, along several dimensions, is an important indicator of successful knowledge exchange. Data are situated in contexts, are malleable, difficult to make mobile, and never are truly 'raw' (Borgman, 2015; Bowker, 2013; Latour, 1987; Loukissas, 2019; Rosenberg, 2013). Many individuals, institutions, and technologies are involved in data exchange processes. Thus, it takes a village to share research data (Borgman & Bourne, 2022).

In our formulation, the farther away a prospective reuser is from the data creator, the more labor that will be required to reuse those data, and the less likely that successful reuse will occur. Imagine a researcher who creates a dataset about energy usage in a region. For climate modelers to reuse these datasets, the best investment may be in software for ease of ingestion by models and detailed metadata descriptions of temporal capture. For teachers to reuse the same energy datasets in teaching about sustainability, the best investment may be in educational documentation. For a policy maker to reuse the dataset, the data creators' best investment may be in matching the data formats and tools to those commonly used in the science policy community.

We turn the question of how to share data upside down, instead asking who might reuse data, how, for what purposes, and when? By comparing relationships between data creators and prospective data reusers along six dimensions of distance, we can scope data creators' roles and responsibilities more precisely. Data creators carry a large burden in making their data useful to others, one that is poorly understood as a form of knowledge exchange. We draw upon the literature of data sharing and reuse, and our own research, to offer principles and practices that can improve knowledge exchange and make data sharing more effective for the many stakeholders in the process (Borgman, 2015; Faniel & Yakel, 2017; Groth et al., 2019; Harper, 2023; Leonelli, 2016; Leonelli & Tempini, 2020; Pasquetto et al., 2017, 2019; Wallis et al., 2013; Wang et al., 2021; Zimmerman, 2007).

We focus on the value created in the relationship between those who create and those who reuse research data. While other kinds of reuse exist, such as data mining, non-consumptive use, 'distant' reuse (as in the sense of 'distant reading'), and bibliometrics, these are outside the scope of our investigation.

The structure of this article is as follows. First, in the remainder of this section, we lay the ground work for developing the construct of distance by discussing the burdens of data sharing and reuse situated within the larger context of knowledge infrastructures. Secondly, in Section 2, we detail the six dimensions of distance between data creators and reusers. Section 3 discusses the implications of employing this construct for stakeholders in the research data ecosystem and for further research. To conclude, we summarize key findings (Section 4.1) and provide practical recommendations (Section 4.2) for data creators, data reusers, data archivists and funding agencies.

### *1.1 Shaping Data Sharing*

Implicit in data sharing policies are assumptions that research datasets are valuable entities worthy of stewardship, are useful to others, and that they will be reused. None of these assumptions are absolute, however. The research enterprise produces far more data than can be kept, whether for short or long periods of time. Some data will prove to be very useful to future researchers; other data may never be reused by anyone.

Multiple actors shape decisions about what data to share, to keep, to store, to discard, and by what practices. Data sharing is rife with ethical concerns for equity, access, and misinterpretation (Lee & Dourish, 2024; Nature Editorial Board, 2021; Staunton et al., 2021). Funding agencies and journals vary in the scope of their mandates for data sharing and in their degree of enforcement. Ethical considerations influence what data fall within the scope of these mandates, and how such data are handled. Human subjects data that are governed by institutional review boards, health records, personally identifiable information, and data that are sensitive for other reasons all tend to receive special handling. Qualitative data such as ethnographies may be treated as sensitive due to the difficulties of anonymization and interpretation (Weller, 2023). Data archives that are certified for handling sensitive data may accept such material and provide controlled access to qualified researchers. Determining what data are worthy of long-term investment is the grist of archival work, and one of continual reassessment by data creators and archivists alike (Acker, 2015, 2017; Baker & Mayernik, 2020; Baker & Yarmey, 2009; Borgman, Golshan, et al., 2016; Faniel & Yakel, 2017; Mayernik, 2016, 2019; Mayernik & Acker, 2018; Staunton et al., 2021; Treloar & Klump, 2019).

The volume, variety, and velocity of data production (Laney, 2001) also influence what can or should be preserved. Researchers who produce complex datasets may process and reprocess those data into many states for analysis. Many instruments, field conditions, software packages, algorithms, protocols, and hands may 'touch' a dataset between the origination of a project and publication of a paper.

Norms for data sharing also vary widely by discipline, domain, and other factors explored throughout this article. Generally speaking, data sharing is more extensive, and more heavily enforced, in the physical and biological sciences than in the humanities. Norms in the social sciences vary between quantitative and qualitative research, sensitivity of data, and community, for example. Temporal factors such as concerns for priority and 'scooping' are addressed in later sections.

### *1.2 Burdens of Data Sharing*

Responsibility for sharing data generally falls to the principal investigators of a research project. Among the many tasks that may be involved are determining the scope and type of data to release, verifying data release requirements of funding agencies and journals, cleaning data, de-identifying human subjects records, writing documentation, describing data with adequate metadata and provenance information, developing training material, registering and obtaining a Digital Object Identifier (DOI) to identify a dataset persistently, and submitting the dataset package to an appropriate archive. Most of these activities are 'soft costs' that are difficult to quantify, and that depend upon labor of highly skilled workers (Hudson Vitale, 2023; Johnston et al., 2018, 2024a; Marsolek et al., 2023a; Merrill, 2014; National Academies of Sciences, Engineering, and Medicine, 2020; Vertesi & Dourish, 2011). However, they are still costs.

Recommendations to data creators for how to invest in their data are plentiful; a recent review identified 35 actions to facilitate data reuse (Koesten et al., 2020). Interview studies and ethnographies on data reuse generally find that researchers need help in producing quality metadata, in addressing creation and usage standards, managing compliance, and more training in research data management (Donaldson & Koepke, 2022; Faniel et al., 2016; Faniel &

Jacobsen, 2010, 2010; Faniel & Yakel, 2017; Mayernik, 2016, 2019; Mayernik & Liapich, 2022).

Once data are released, maintaining access to them entails further expenses that fall upon data creators, their institutions, data archives, or other parties. These costs include computational resources to host the datasets, staffing to ensure that the resources remain operational, continuing curation to migrate datasets as underlying software changes, and curatorial staff to assist in deposit, search, retrieval, and reuse of datasets.

Policy makers are increasingly concerned about the costs imposed by data sharing requirements. The U.S. National Library of Medicine commissioned the National Academies of Science to conduct a cost study of biomedical data production and stewardship. Their extensive report identified 44 cost drivers across 21 activities needed to create, maintain, and preserve biomedical data resources. While the Committee provided templates for assessing costs, including salary ranges for the kinds of labor involved, they did not attempt to establish precise costs by type of data or reuse (National Academies of Sciences, Engineering, and Medicine, 2020). In Europe, the Dutch Digital Heritage Network studied costs associated with making cultural heritage data available across their 12 member institutions, focusing on infrastructure for data archiving. Labor and staffing constituted 72% of total costs, including work to ingest, select, and add metadata (Uffen & Kinkel, 2019). Because the scope and type of investments needed to make datasets reusable by others varies widely across domains and contexts, actual costs are very difficult to determine.

Box 1 contains examples of concrete cost estimates. Much of these costs involve ‘invisible work’ of the sort that becomes visible only when something breaks down because the work is not done (Borgman, 2003; Crain et al., 2016; Star & Ruhleder, 1996).

**Box 1 about here**

Box 1: Examples of costs and outcomes of data sharing

1. The Alzheimer’s Disease Neuroimaging Initiative (ADNI) provides clinical, imaging and biomarker data about the progression of Alzheimer’s disease. A decade ago, when the annual project cost was roughly $130 million, about 10-15% was spent on data sharing activities (Wilhelm et al., 2014). Per a more recent estimate, datasets hosted by ADNI have yielded more than 3500 publications from 140 million or so downloads (Veitch et al., 2022). The ADNI data also are used to improve image recognition systems.
2. *Our World in Data* is a popular resource for journalists and educators, with datasets on themes ranging from population statistics to sustainability, at a cost of about $1.8 million per year[1] to maintain their data, metadata, contextual articles, and infrastructure (*Our World in Data*, 2024). Recent estimates on their website indicate that their COVID-19 data is cited in 5000+ papers and $CO_2$ emissions data in 2500+ papers. .
3. *Common Crawl*, launched in 2007, now spends roughly $200,000 per year to maintain petabytes of openly available web crawl data (*Common Crawl*, 2023; Roberts, 2013). The use of these data to train large language models has dramatically changed the course of AI research and practice (Baack, 2024; Min et al., 2024).

**End of Box 1

---

[1] Computed based on costs reported in the 2022 annual report of the Global Change Data Lab (parent organization of *Our World in Data*).

### *1.3 Burdens of Data Reuse*

Data reusers also incur costs to acquire, process, analyze, interpret, and redeploy datasets. We provide numerous examples of barriers to reuse that are imposed by data sharing practices, however unintentional.

Some researchers are net data creators and rarely reuse others' data. Other researchers are net reusers, rarely creating their own data resources through observation, experimentation, or other methods. Those researchers who are both data creators and reusers are likely more sensitive to the nuances involved in knowledge exchange. A researcher may acquire a dataset, alter it by some means such as further processing, addition or removal of data points, supplementary documentation, or migration to other technical platforms, and then share the new version of the dataset. Each dataset iteration released may create new burdens for reusers and for data archives (Gregory et al., 2020).

### *1.4 Knowledge Infrastructures*

Our aim here is not to provide a comprehensive review of the literature on data sharing and reuse. Rather, we draw on that literature and our own research to characterize distances between data creators and reusers. As research data exist in a web of complex social and technical relationships, we develop the distance construct in the context of knowledge infrastructures – defined as 'robust networks of people, artifacts, and institutions that generate, share, and maintain specific knowledge about the human and natural worlds' (Edwards, 2010, p. 17).

We also acknowledge that 'data,' even when constrained to research contexts, remains a contested concept. The term 'data' has evolved in uses and meaning over the course of scientific history, originating in mathematics before expanding into other areas of the physical and life sciences (Meyns, 2019). While 'data' are now associated with notions of evidence and facts, neither has been the case throughout history (Blair, 2010; Daston, 2017; Meyns, 2019; Puschmann & Burgess, 2014; Rosenberg, 2013, 2018).

Acts of 'creating' data involve many kinds of expertise and many steps in judgment and selection. Data may originate as observations; outputs of experimental apparatus; be generated by computational models; collected as samples, specimens, or artifacts; or obtained by other methods specific to a research domain or topic of study. Many people may be involved in creating an individual dataset, each contributing expertise and making decisions about what to capture or trust (Baker & Mayernik, 2020; Mayernik, 2016; P. Zhang et al., 2023). Some, but by no means all, of these decisions can be documented and associated with a dataset. As a consequence, the originators of a dataset retain the most intimate knowledge of its content (Pasquetto et al., 2019). .

Knowledge infrastructures can shorten or extend distances between data creators and reusers. Some of the expertise involved in data creation and reuse is close at hand, in the form of research team members, laboratory technologists, software developers, data managers, librarians, archivists, and other specialists participating in research projects. Other entities influencing data exchange are farther away, and perhaps invisible, such as institutional infrastructure for the research enterprise; developers of software, hardware, and other technologies; maintenance and repair units; generic and domain-specific data archives, standards agencies, publishers, editors, and various domain-specific stakeholders (Aspesi & Brand, 2020; Bierer et al., 2017; Bly & Brand, 2023; Borgman & Brand, 2024; Brand, 2022; Castro et al., 2017; Eschenfelder et al., 2022; Howison & Bullard, 2016; Katz et al., 2021; Katz & Hong, 2024; *Maintainers*, 2024;

McKiernan et al., 2023; Russell & Vinsel, 2018; The Information Maintainers et al., 2019; Thomer et al., 2018; Weber et al., 2012; P. Zhang et al., 2023).

## 2 Dimensions of Distance between Data Creators and Data Reusers

We develop the construct of *distance* between the data creator and the potential reuser as a model for knowledge exchange. This framing was originally proposed in Borgman (2015) and complements several other socio-technical models for data practices. We use the construct of distance as a metaphor in the noun sense of 'the extent or amount of space between any two points; the length of space that must be traversed in order to get from one place or thing to another' ("Distance," 2024). Because metaphors are useful to comprehend social and scientific phenomena, they are widely studied in fields such as philosophy, sociology, and social studies of science (Lakoff & Johnson, 2003; Mager & Katzenbach, 2021; Puschmann & Burgess, 2014; Wyatt, 2021). As Wyatt (2021, p. 1701) explains in the context of critical Internet studies, metaphors are useful to make complex phenomena meaningful and to shape the meaning of phenomena. Baker and Yarmey (2009) similarly use the phrase 'distance from origin' metaphorically in reference to data repositories, describing how close or far they operate from the sources of data they acquire. Leonelli and others use the metaphor of 'data journeys' to explore the lineage of data as they travel across contexts over time (J. Bates et al., 2016; Leonelli, 2016; Leonelli & Tempini, 2020). Metaphors abound for 'data' as oil, fuel, gold, or other substances (Puschmann & Burgess, 2014); and for the Internet as a library, a medium for communication, a market, a digital world, or a highway (Borgman, 2000; Wyatt, 2021). Likewise, Lane and Potok's (2024) vision for 'democratizing data,' as a special issue theme in this journal, addresses the need for infrastructures to facilitate data access for a wider array of communities (Borgman & Brand, 2024; Christen & Schnell, 2024; Nagaraj, 2024) .

Our theory of distance between data creators and those would use those data focuses on questions of how, why, for whom, and for how long data are reusable. Given that data creators cannot anticipate all possible reuses or reusers, our aim is to identify factors that would aid data creators and other stakeholders in deciding how to invest in their data, how to share them, and how to identify potential reuses and reusers.

We identify six dimensions of distance: domain, methods, collaboration, curation, purposes, and time and temporality. These distance dimensions are primarily social in nature, involving individual and community practices and norms, and relationships with knowledge infrastructures. Technologies play important roles in knowledge exchange, both to 'lubricate' the transfer of data between creators and reusers, and to cause 'friction' between data creators and prospective reusers (Edwards et al., 2011). Hence, we illustrate the socio-technical nature of the distance dimensions by examining particular technologies associated with each dimension. While we attempt to characterize each distance dimension as distinctly as possible, we also acknowledge their interdependencies.

Data sharing and reuse involve multiple actors, with multiple relationships. Our distance model is concerned primarily with data creators, data reusers, and data archivists as mediators, while acknowledging that many other actors and infrastructure components are in play. As illustrated by examples throughout this section, data creators and reusers may interact directly or indirectly. Direct exchange occurs when creators transfer datasets to prospective reusers, such as collaborators, students, or requestors – intentionally and in real time. More commonly, reusers acquire datasets indirectly, such as via a data archive or website maintained by the data creators. Direct exchange offers the greatest opportunity for interpersonal communication about the datasets. Indirect exchange introduces more distance between creators and reusers, and more actors in between them.

### *2.1 Domain Distance*

Studies of data sharing and reuse commonly focus on a domain of some sort, whether community, expertise, epistemology, technology, language, or other grouping. The term 'domain' is frequently taken at face value, rather than explored as complex construct with subtle influences on data practices (Borgman, 2015; Koesten et al., 2020). Domain distances often cascade through the other five dimensions, hence we frame it first.

Research data are created within domains of 'human action or expertise' (Ribes et al., 2019, p. 283). Domains might be defined as broadly as art history or as narrowly as the community that employs a particular method to handle genetic material associated with craniofacial abnormalities in zebrafish. Data creators and reusers with common expertise in theory and method will find data exchange easier than those with contrasting expertise.

The 'logic of domains' is best understood as an alternative to generic, cross-domain, or domain-independent data collections, methods, or tools (Ribes et al., 2019). Domains address common knowledge and purpose, whether for data collections, technologies, or reasoning in artificial intelligence. Scholars in fields as diverse as philosophy, economics, education, and software engineering wrestle with the difficulties of drawing boundaries around academic disciplines. Domain boundaries are necessary, even if drawn arbitrarily, to demarcate academic departments, funding agency programs, scholarly conferences, library collections, data collections, and communities.

As discussed in more depth elsewhere (Pasquetto et al., 2019), differences in knowledge and expertise help to explain the reusability of research data. A useful distinction, drawn from epistemology, is between 'knowledge that' and 'knowledge how' (Ryle, 1949). 'Knowledge that' something exists may be sufficient to locate a document or dataset of interest, but more advanced expertise in 'knowledge how' something works may be necessary to interpret and reuse those data. Degrees and types of tacit knowledge can influence the ability to interpret and reuse data (Collins et al., 2007; Collins & Evans, 2002, 2007; Galison, 1997; Schmidt, 2012).

Domain boundaries appear throughout the life cycle of data creation, sharing, reuse, and stewardship, whether explicitly or implicitly. The Open Archival Information System (OAIS) reference model, which is widely deployed for data archives, refers to the 'designated community' – *An identified group of potential Consumers who should be able to understand a particular set of information'* – served by the data archive (Consultative Committee for Space Data Systems, 2012, pp. 1–11).

Infrastructures for research data are typically organized by and for domains, whether by funding agencies, professional societies, or other sets of stakeholders (Bastow & Leonelli, 2010; Borgman, 2007; Edwards, 2010; Star & Griesemer, 1989; Star & Ruhleder, 1996). Domain-specific infrastructure facilitates data exchange for those who have access to the shared infrastructure, thus decreasing distance between data creators and reusers.

Astronomy offers an illustrative case study of a large, long-term, loosely coordinated, international, knowledge infrastructure funded by numerous agencies. Most major telescope observatories host data archives, alone or in partnership with a funding agency, as discussed further below under *Curation Distance*. NASA supports the Astrophysics Data System, managed by the Harvard-Smithsonian Center for Astrophysics, which hosts bibliographic records of publications in astrophysics and some related fields (NASA/ADS, 2024). The Centre de Donnees astronomiques de Strasbourg (CDS) in France manages and hosts metadata catalogs of named astronomical objects (Genova, 2013, 2018). These are but a few of the many infrastructure components that link directly to each other, supporting intensive international data traffic. As a consequence of these and many other infrastructure investments, data exchange in astrophysics is easier than in most other domains, and is now standard practice in the field (Borgman, Darch, et al., 2016; Borgman & Wofford, 2021). The success of domain-specific infrastructures can create barriers for cross-domain research, and thus distance, between data creators and reusers who rely on incompatible infrastructures.

Figure 1 suggests relative domain distances between data creators and reusers. In this example, an astronomer specializing in infrared spectra is positioned a short distance from other infrared astronomers, whom they may view as 'scholarly siblings.' At the other extreme are prospective reusers such as artists or musicians who wish to employ astronomy data in their creative activities, who may be viewed as 'intellectual strangers' relative to the astronomers.

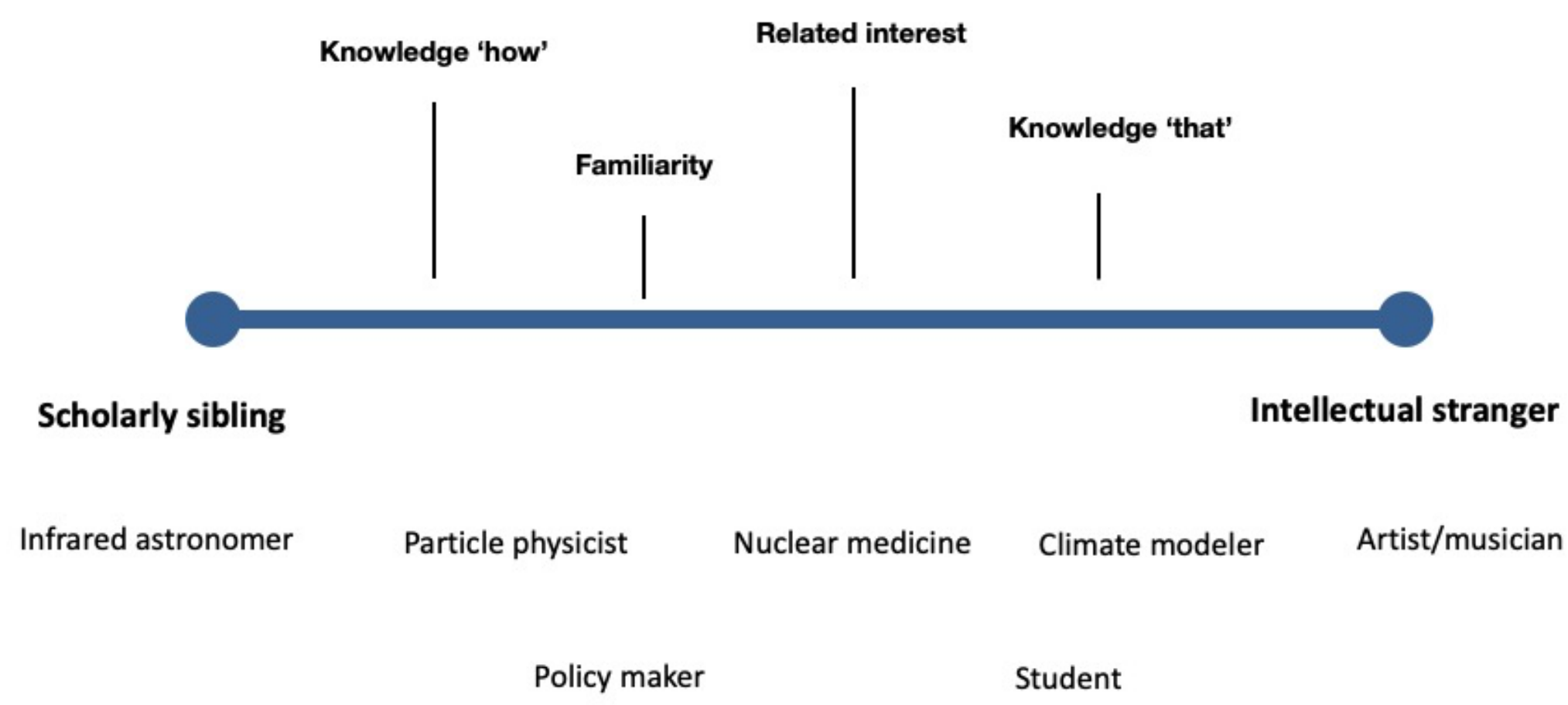

Figure 1: Example of relative distance in domain from an infrared astronomer to categories of potential data reusers.

In sum, the more shared expertise in a research domain between data creator and prospective reuser, the less distance between them, and the greater likelihood of knowledge exchange.

### *2.2 Methods Distance*

Researchers' choices of methods are frequently a function of disciplinary expertise and available infrastructure in the university, laboratory, collaboration, community, or other context. Among the many methodological factors that may influence the compatibility of data creation and reuse are quantitative vs. qualitative, laboratory vs. field, human interaction vs. artifactual, size and scale, short vs. long term, analog vs. digital, hand-collected vs. machine-generated, degree of sensitive information, and choices of language, whether linguistic or programming.

Scale factors also influence methods options. To produce gigabytes or terabytes of data, whether in physics, climate science, or humanities, machine-generation is required. Conversely, those who conduct fieldwork, whether in ecology, ethnography, or earth sciences, may acquire small volumes of hand-crafted data that must be described and documented manually. The specific details of any of these methods may vary widely within and between research domains.

As graduate students, new scholars learn suites of research methods and may develop new ones. Some researchers spend their careers within a narrow set of methods and instruments, while others apply methods across domains, or multiple methods within a domain. As scholars collaborate with colleagues, students, and other partners, their methods may complement or converge (Darch, 2016), as discussed further in the next section. The ability to interpret data produced by others may depend heavily upon knowledge of the research methods by which those data were created. Researchers who are expert in experimental designs may have difficulty interpreting survey data, and vice versa. Those who work with one model organism, such as a certain type of laboratory mice, may not be familiar with methods associated with other model organisms, such as zebrafish (Pasquetto, 2018). Similarly, deep expertise in exploiting a particular telescopic instrument affords an astronomer scientific advantage over those new to the methodological environment.

Emergent research topics are sources of innovation in methods, as scholars seek new ways to ask new questions. Innovative methods may require 'magic hands,' whereas data exchange in areas where data and tools become 'kits' is common practice (Hilgartner & Brandt-Rauf, 1994). New methods necessarily lack common standards and thus limit data exchange. Darch and others studied a large long-term infrastructure development in the deep subseafloor biosphere, comparing research topics pursued and scientific methods employed (Darch et al., 2015; Darch & Borgman, 2016). Much of the shipboard technology and data handling facilities were constructed decades earlier to support physical science research. The physical scientists had standardized their methods and data practices to facilitate data exchange within their communities. The biological scientists studied by Darch et al. were exploring new questions, using new research methods. Partly as a consequence of the lack of standardization, the biologists' data handling methods were more labor-intensive and exchange of data between collaborators was more complex. The biological scientists were engaged in a debate over when, why, and how to standardize their methods. Some argued that standardization was premature, as it would constrain the development of a new area of science. Others argued that standardizing methods would enable them to become a 'big science' sooner, grow their community, and obtain more funding.

While data production at large scales is more automated than artisanal data collection at smaller scales, machine-generated data and metadata do not necessarily result in datasets that are easily exchanged. Datasets may be standardized to the specifications of the machines' manufacturer and vary by model, for example. Similarly, automatically generated metadata may be interoperable only to the extent that machines were programmed with data exchange in mind. Complex instrumentation for the physical, biological, and life sciences relies on software pipelines that clean and process data to make them scientifically viable, yet each instrument may have multiple pipelines tailored to the methods employed by individual researchers (Bowers et al., 2006; Brinckman et al., 2019; David et al., 2023; Li & Ludäscher, 2023; McPhillips et al., 2009; Yehudi et al., 2023).

While standards are an embodiment of infrastructure (Star, 1999; Star & Ruhleder, 1996), they range in implementation from formal, legally enforceable standards (de jure) to informal standards (de facto) that are employed by community consensus (Lampland & Star, 2009; *National Information Standards Organization*, 2024). De facto standards are commonly employed for data exchange, methods, or protocols, whereas some fields do establish de jure standards. In practice, individual research teams may 'standardize' their own data production in ways that are internally consistent, but not necessarily interoperable with those of other teams.

The undersea biosphere case echoes earlier scientific debates about the risks of premature standardization and the benefits of standardization to advance a field (Leonelli & Ankeny, 2012). These debates continue, addressing issues such as the benefits and risks of 'pre-registering' studies for the purposes of replication (Chen et al., 2018; Rocca-Serra et al., 2023; Serghiou et al., 2023). Social psychology, for example, promotes the use of registered reports, which is a mechanism to make public the protocols for collecting and analyzing data in advance of conducting research (Alipourfard et al., 2021; Nosek & Lakens, 2014). In response to concerns that registered reports enable others to 'scoop' research, some sites now allow researchers to hide their registrations until their findings are published (Open Science Foundation, 2024).

In sum, data reusers who have the most intimate familiarity with the methods by which a dataset was created will have the shortest distance from origin, and thus likely an easier time reusing data than researchers with minimal expertise of those data creation methods.

### *2.3 Collaboration Distance*

Scholars collaborate but they also compete with each other. Cooperative practices emerged in the 17th century with the first scholarly journals, when scholars and scientists saw benefits in working together (Daston, 2023). Scholars may find ways to collaborate in some areas, such as sharing data and employing common technologies, while still competing for grants, prizes, jobs, and other rewards (Borgman, 2007). Rates of co-publication, as an indicator of collaboration, have increased significantly in recent decades, in parallel with increases in specialization (Frenken et al., 2009). Although geographic distance between collaborators is less of a barrier in an age of air travel and networked communication technologies, physical proximity still matters. Early studies of distributed collaboration established the value of co-location to establish and nurture relationships (Bos et al., 2008; Jirotka et al., 2013; Olson et al., 2008; Olson & Olson, 2000; Q. P. Zhang et al., 2004). Scholars became adept at online collaboration in the isolation period of the COVID-19 pandemic, while placing great value on face-to-face interaction as restrictions eased.

In principle, collaborators should be able to exchange data easily, whereas strangers should have more difficulty in interpreting and reusing datasets. In practice, collaborations take time to mature, as researchers and teams find common ground across their domain knowledge, methodological expertise, and scholarly practices (Bowker, 2005; Gebel et al., 2024; Jirotka et al., 2013; Olson et al., 2008; Olson & Olson, 2000; Ribes & Bowker, 2008).

'Collaboration' may refer to a formal agreement to work together, such as joint funding or organizational structure, but it can also refer to less formal collegial relationships among scholars who work together intermittently. Over the course of a career, scholars develop 'invisible colleges' (Crane, 1972) of people with whom they collaborate, exchange research papers, organize conferences, invite to speak, and may exchange data. Academic lineage is a feature of invisible colleges: advisors may exchange data more readily with their current and former students than with scholars with whom they have more distant relationships. Knowledge is situated in communities, however difficult it may be to draw boundaries between them. Social science theorizing about knowledge and communities also falls under rubrics of 'communities of practice' (Lave & Wenger, 1991) and 'epistemic cultures' (Knorr-Cetina, 1999). Individuals may belong to multiple communities concurrently, each of which evolves over time.

Datasets often serve as 'boundary objects,' a construct used in science and technology studies to describe entities that lie at the intersection between communities (Star & Griesemer, 1989). Boundary objects such as datasets, methods, and terminology may be understood differently by each partner in a collaboration. Collaborators necessarily translate, negotiate, and simplify meaning of such boundary objects to facilitate knowledge exchange.

While many social factors, practices, and technologies can lubricate collaborations or cause friction (Jackson et al., 2010), data formats and associated software tools are common barriers to data exchange. Data friction arises when collaborators, or other reusers, encounter data formats for which they lack the tools to open, much less interpret, manipulate, and process a dataset (Edwards et al., 2011). As noted above, datasets often are output from commercial hardware or software in proprietary formats. Such files may be readable only within the same proprietary setup, including the same versions of hardware and software. If such datasets are shared with collaborators or submitted to repositories with sufficient documentation that others can recreate the environment, the datasets may be reusable or might be transferable to other environments. However, the labor and cost of setting up the technical environment to reuse a given dataset may be infeasible.

A common means to reduce collaboration distance is to rely on 'lowest common denominator' formats such as CSV, Excel, or Google Sheets that can be exchanged. In 2020, 37% of the data indexed by Google was in tabular formats such as CSV or XLS; another 30% was in structured formats such as XML or JSON (Benjelloun et al., 2020). When collaborators with different data format and tool environments wish to exchange data, they may resort to lowest common denominator formats, despite the information loss involved. This situation occurred often in the Center for Embedded Networked Sensing, a decade-long NSF center (2002-2013). Some teams relied on Matlab for data analysis, while others employed R for statistical processing. When necessary to merge, exchange, or compare data, collaborating groups would export the relevant data to Excel spreadsheets (Borgman, Golshan, et al., 2016).

While useful for data exchange, common denominator formats rarely allow users to structure, format, annotate, and add metadata to datasets. These exchange formats limit functionality for subsequent data analysis and for interoperability across technical environments.

In sum, because collaborators have incentives to share data and to explain research contexts to each other, the relationship provides a means to shorten distances between data creators and data reusers.

### *2.4 Curation distance*

Curation is the process of adding value to artifacts, whether data, documents, museum objects, or other entities worthy of stewardship (Marsolek et al., 2023b). Data creators curate their data by verifying, describing, documenting, and registering datasets. Archivists, librarians, and museum curators manage artifacts to the standards and practices of their fields and institutions. The degree of curation given to any dataset varies widely, from minimal description to elaborate care and maintenance. Many (or few) hands, processes, and institutions may touch a data product over its lifetime (Mayernik & Acker, 2018; Parmiggiani et al., 2023). The type, amount, and relevance of curation a dataset receives influences the distance between data creators and reusers. Journal articles are the primary form of documentation for many datasets, providing explanations of research methods, hardware and software, and necessary detail about technical environments, such as software versions and parameters, that are required to open files and execute routines (Bourne, 2005).

Metadata, as curation practice and as a form of documentation, may increase or decrease any of the six dimensions of distance between data creators and reusers. Loosely defined as 'data about data,' metadata are essential to the exchange of data products. While the term is often used as a simple shorthand for any description of information artifacts, from books to datasets, metadata has a rich intellectual history in knowledge organization (Getty Research Institute, 2008; Gilliland-Swetland, 2000; Mayernik, 2019, 2020a; Mayernik & Acker, 2018; Zimmerman, 2007). Professional practice for creating metadata varies widely across the 'memory institutions' of libraries, archives, and museums. Among the categories of metadata identified by Mayernik (2020b) in a review of professional curatorial practice are these: access, administrative, archive, authentication, browse, character, descriptive, discovery, finding, identification, linking, preservation, provenance, relationships, rights, structural, technical, understanding, and use.

Research data archives, which serve as important mediators in knowledge exchange, vary greatly in content, in maturity, scope of community, funding sources, degree of stewardship commitment, and many other factors. A simple three-level categorization introduced by the U.S. National Science Board (NSB) in 2005 remains useful in discussing curation distance. *Research data collections* originate from research projects on narrow topics, receive minimal processing, and may be valuable for specific purposes. Those collections with longer term value and support may establish or employ community standards, becoming *resource* or *community collections.* The third category, and the one usually implied by the term 'data archive,' are *reference data collections*. The latter have long-term stewardship commitments, sustainable funding, and conform to trust and technical standards (Borgman, 2015; National Science Board, 2005).

Those research data archives with highest levels of commitment to stewardship subscribe to community standards such as the TRUST (Transparency, Responsibility, User focus, Sustainability, and Technology) principles for digital repositories (Lin et al., 2020). These principles, in turn, promote standards for digital stewardship such as the Open Archival Information System (OAIS) reference model and certification such as the CoreTrustSeal (Consultative Committee for Space Data Systems, 2012; *CoreTrustSeal*, 2024). At the other extreme are datasets released without documentation, such as cases in which data creators post links to datasets that simply trigger downloads of spreadsheets or other files (Mandell, 2012; Wallis et al., 2010).

Data quality guidelines consistently emphasize the need for comprehensive documentation including rich descriptive metadata, useful human-readable and machine-readable descriptions of the dataset, links to background information, and descriptions of attributes (Publications Office of the European Union, 2021; Van de Sompel & Nelson, 2015; Wilkinson et al., 2016). Baker and Mayernik (2020) describe in detail the workflow processes by which archivists add value to datasets contributed to repositories and how those activities differ from the work of scientific data creators. These curation activities are labor-intensive and require extensive consultation between data contributors and data creators. As a consequence, curated datasets may be released weeks or months after initially submitted to an archive.

Some research domains make large infrastructure investments in data archives. Biomedical data archives maintained by national and international agencies are important sources of genome sequences, for example. Space-based astronomical observatories such as the Hubble Space Telescope, Chandra, James Webb Space Telescope, and Gaia include data archiving as a core component of their scientific missions. However, support for data stewardship varies even within data-intensive domains such as astronomy. By contrast to space-based missions, long-term support for data stewardship was not included in initial mission funding for ground-based astronomical observatories such as the Sloan Digital Sky Survey and the W. M. Keck Telescopes (Borgman, Darch, et al., 2016; Borgman & Wofford, 2021; Darch et al., 2020, 2021).

Data stewardship also varies widely in the social sciences and humanities. Social science archives with long-term government funding for data stewardship include GESIS in Germany, DANS in the Netherlands, and the UK Data Archive. ICPSR, a US-based social sciences data archive, has served a broad international community for more than 50 years as a membership organization (Data Archiving and Networked Services, 2024; Interuniversity Consortium for Political and Social Research, 2024; Leibniz Institute for the Social Sciences, 2024; *UK Data Archive*, 2024). Data archives can shorten the distance between data creator and data reuser through their curatorial processes, both in data handling and by direct assistance. As explored in a case study of the Digital Archiving and Networked Services of the Netherlands (DANS), archivists add value to data exchange in several ways. They work with creators who are prospective data depositors to describe, document, and validate the contents of datasets. These activities make datasets more discoverable and usable. Similarly, archivists work with prospective data reusers to aid in searching, retrieving, interpreting, and using datasets in their repository (Borgman et al., 2019; Clement & Acker, 2019; Jung et al., 2022). A study of 1,480 users of ICPSR data also showed that quality documentation resulted in higher levels of satisfaction (Faniel et al., 2016).

By contrast, self-curated repositories can reduce the distance between data creators and reusers by improving the discoverability and retrievability of datasets, but to a lesser degree than domain-specific archives that invest heavily in curatorial services. Self-curated data repositories vary in the amount of automated verification applied to content, in the diversity of types of material they contain, and the degree to which they build a coherent collection for a research domain or topic. Primary advantages of repositories such as Zenodo, Figshare, and university digital archives are cost, speed, and scale (Asok et al., 2024; *Figshare*, 2024; *Zenodo*, 2024; Johnston et al., 2024b; Kruithof et al., 2024; Ndegwa & Kuchma, 2024; Ott et al., 2024; Sofi et al., 2024; Verburg et al., 2024). These repositories are akin to the NSB categories of research and resource data collections (National Science Board, 2005). DataVerse, which originated as a social sciences data repository at Harvard, is also a software platform that is widely adopted by universities, research institutes, and data archives for data repositories (*The Dataverse Project*, 2024). Data creators bear sole responsibility for documentation in self-curated repositories such as Zenodo or Figshare, whereas heavily curated repositories such as ICPSR or DANS usually provide more extensive descriptions and additional metadata for the datasets they acquire.

Among the many lessons learned from arXiv in 30 years of sharing preprints in physics, computing, and related fields is that 'minimalist quality control' via machine learning methods can support the 'unforgiving daily turnaround' of ingesting papers at scale. ArXiv employs mechanical means of inspection to flag problematic submissions for humans to review, avoiding some of the worst cases of inappropriate or hazardous postings (Ginsparg, 2021). In studies of data on Github, quality of documentation is also a predictive factor for data reuse (Koesten et al., 2020).

While research data discovery is improving with the development of search engines designed specifically for datasets, retrieval still relies heavily on the quality of documentation (Chapman et al., 2020). Dataset search engines improve with richer metadata (Brickley et al., 2019). When data search engines cannot identify text, links, or other parameters to index, most can only assess the metadata of datasets (Gregory & Koesten, 2022). Some data quality checks can be performed at the ingest stage (Servilla et al., 2016). Experimental dataset search engines can probe datasets at the time of retrieval to generate basic profiles and then extract information to create metadata such as types of the columns used and dataset summaries (Castelo et al., 2021). Other advanced systems rely on the webpages associated with a dataset for focused crawling and dataset discovery (H. Zhang et al., 2021).

A perpetual problem with searching for data is the volume and variety of potentially useful data points and observations that are locked up in tables, figures, and narrative of PDF documents such as journal articles and reports. While text in newer PDF formats is searchable, the lack of markup for tables and figures often requires the prospective data reuser to extract datapoints manually (Choi & Xin, 2021; Lawrence, 2022).

Users of dataset search engines frequently are frustrated by the difficulty of discovering datasets of interest, due to lack of searchable documentation, user interfaces that lack pertinent parameters, and other search engine limitations (Gregory et al., 2019, 2020; Koesten et al., 2021). When available datasets are camouflaged by these searching constraints, distance between data creators and reusers increases.

Interoperability between retrieval systems, collections, and metadata has been a holy grail since the earliest days of digital library research (Directorate-General for Research and Innovation, European Commission et al., 2021; Lagoze, 2001; Lynch & Garcia-Molina, 1995; National Institutes of Health, 2015; Nyberg Åkerström et al., 2024; Sansone & Rocca-Serra, 2016; Van de Sompel et al., 2006; Van de Sompel & Nelson, 2015). Contemporary efforts at interoperability, especially those driven by the FAIR (Findable, Accessible, Interoperable, and Reusable) principles (Wilkinson et al., 2016), focus on 'semantic interoperability' between systems with the goal of enabling machines to exchange data with less human intervention in the search process (Directorate-General for Research and Innovation, European Commission et al., 2021; Nyberg Åkerström et al., 2024).

In sum, the more curation applied to datasets, and the more trusted the data archive, the less distance between the creators and prospective reusers. The greatest curation distance occurs when datasets are posted or linked with minimal documentation or validation.

### *2.5 Distance in Purposes*

Research data are not disembodied natural objects that are readily repurposed. Rather, they are created in specific contexts to serve specific purposes and usually to address specific questions. The embeddedness of data in contexts is a topic well studied in philosophy, social studies of science, history, and within scientific domains. Also common to findings about the embeddedness of data are the difficulties of making data 'travel' beyond their origins. Considerable information loss occurs when data are removed from their original contexts (Bowker, 2005; Latour, 1987, 1993; Latour & Woolgar, 1979; Leonelli & Tempini, 2020; Loukissas, 2019).

Contexts for data creation and reuse vary by all the factors noted above, such as domain, discipline, community, team, methods, and laboratory or field conditions. When a prospective data reuser wishes to verify or reproduce a study that is described in a journal article, contexts may be sufficiently similar that distance is short. When a prospective reuser seeks to employ datasets for very different purposes, such as the above example of using a dataset on local energy usage for environmental modeling, teaching, or science policy analysis, the distances between creators and reusers will be much greater.

In a study of data reuse in biomedicine and environmental research, Pasquetto et al. (2019), found strong differences in ease of reuse between comparative and innovative purposes. Researchers frequently reused data from archives and repositories to compare or 'ground-truth' their own studies. Such reuses of available datasets were so common that neither the datasets nor the archives from which they were acquired tended to be mentioned or cited in publications. Only in rare cases did researchers seek to reuse datasets to ask innovative new research questions, thus, to reuse data for purposes other than those for which the dataset was originally created. In these repurposing situations, the data reusers rarely found the contextual documentation sufficient. Rather, they contacted the data creators directly for more information, and usually initiated a new collaboration for the purposes of reusing and integrating their data.

Baker, Duerr, and Parsons (2016) provide a rare deep dive into how datasets, other data products, and data archives may need to be adapted to new purposes and new audiences over time. The U.S. National Snow and Ice Data Center (NSIDC) was designed originally to serve scientists within a narrow set of domains who were collecting and analyzing snow and ice data. As the value of these data became apparent for emergent questions about climate change, and for a much broader array of scientists and scholars, the NSIDC invested in new ontologies and metadata, and extensive recataloging of older data to serve new purposes. These were expensive investments by the NSIDC and its parent funding agencies, justifiable by the importance of the scientific questions to be addressed.

Other cases exist in citizen science, such as the Zooniverse (2024) project on 'old weather,' in which participants transcribe hand-written weather descriptions from sources such as 19th century ships' logs. Personal diaries from earlier centuries that record dates when flowers bud or bloom, when birds arrive, and other observations of the natural world also are proving useful in assessing climate change. Repurposing old records is useful in certain circumstances, but does not scale. In most cases, those who wish to reuse for new purposes must rely on available documentation and on access to the data creators, where possible. As discussed above under *Curation Distance*, the missions of data archives are framed in terms of serving their 'designated community,' with an acknowledgement that the community may change over time.

Data reusers frequently encounter difficulties in identifying, retrieving, and interpreting datasets created for other purposes. One technical means to bridge these gaps are software interfaces that enable two or more programs to work together. Most are known as Application Programming Interfaces (API). These interfaces can reside on a familiar platform, such as MacOS, Windows, Linux, Apple iOS, or Android, to provide dataset views and analytical tools for data in repositories. APIs thus may expose parts of a dataset, both guiding and constraining data exploration. While not a substitute for quality dataset documentation, APIs may be especially useful for those performing cross-disciplinary research (Jia et al., 2022).

APIs also are useful for curated reference collections that serve broad communities and whose data resources may be of interest to larger audiences. For example, ProteomicsDB, a provider of mass spectrometry-based proteomics data, provides multiple interfaces to explore, render, and identify proteins and their expression in organisms. They offer a comprehensive API that allows connections with other data providers and tools to perform predictive analytics (Lautenbacher et al., 2022). ProteomicsDB deployed new interfaces as means of meeting the FAIR principles (Wilkinson et al., 2016).

In sum, investments in creating and curating data should consider the range of purposes for which these data might be reused by others. Similarly, choices of how and where to archive data will influence the the discoverability of data and the resulting distances between data creators and reusers.

### *2.6 Distances in Time and Temporality*

Whereas the distinction between time and temporality is a basic principle of classical philosophy (Drucker, 2009), these topics rarely are addressed explicitly in data creation, sharing, and reuse. Time and timelines are usually linear and unidirectional, whereas temporality is relational. Events can occur before, after, during, or concurrent with other events, and may be discontinuous or iterative. Data processing involves many time-related factors that may influence the social and technical distances between creators and reusers.

Temporal factors pervade data creation and reuse, both implicitly and explicitly. The most obvious distance aspect is time between when a dataset is created and the time when a prospective reuser wishes to employ those data in later research. Leonelli (2018) distinguishes explicitly between "data time," as the time at which data were collected and analyzed, and "phenomena time," as the duration of time over which those data remain useful as evidence. Months, years, or centuries may pass between data creation and reuse. Practices and circumstances evolve. In the case of digital data, hardware and software ecosystems degrade. If the elapsed time between origin and reuse is relatively short, the parties may readily exchange knowledge about the dataset. If the elapsed time is relatively long, reusers must rely on available documentation and context. Some datasets become more valuable over time, while others decay in value rapidly. Phenomena time, in Leonelli's framing, encompasses all of these factors. As she and others have argued, phenomena lack an intrinsic time-scale. Rather, the useful time period for data is a function of how they are captured, maintained, and applicable to future contexts (Bowker, 2005; Griesemer, 2020; Leonelli, 2018; Wylie, 2024).

To reuse a dataset, a researcher usually needs to know how, why, where, and when it was collected. 'When' can be as simple as a time stamp, but even a time stamp is relative to local or reference clocks, or may be a sequence in a workflow. For example, data pipelines from telescopes normally add timestamp metadata to observations based on Coordinated Universal Time (UTC), which is the successor to Greenwich Mean Time, regardless of the time zone in which the telescope is located. Astronomers plan their observing on local time and date for staffing and conducting remote observing. While UTC and GMT do not observe daylight savings time, local clocks for the astronomers and the telescope may change twice yearly. Astronomers in California, using telescopes in Hawaii, maintain their observing schedule on three time zones that span 10 hours and usually two calendar dates. Thus, 'observing tonight' may begin after midnight local time, and be timestamped tomorrow. When combining observations and reporting results, analysts must be careful to use UTC coordinates.

Scholars in all fields, from artificial intelligence to ancient history, encounter temporal factors that influence how data are collected and interpreted. Loukassis (2019, pp. 74–81) provides an 8-page table of several hundred date formats encountered in processing bibliographic records from the Digital Public Library of America. Archivists wrestle with the many indicators of space and time in records, and with how the meaning and use of records evolves accordingly (Acker, 2017).

Our relational and infrastructural views of data raise questions of data governance, thus, who decides how data should be managed over the long term, and by what criteria – a concern that extends well beyond research data (Davidson et al., 2023; O'Hara et al., 2021). Drucker (2009) develops 'temporal modeling' in the context of the humanities, albeit with a broad framing that is relevant to many areas of scholarship. Representations of time must accommodate 'retrospective, simultaneous, and crosscut temporalities' (Drucker, 2009, p. 37). Allen and Ferguson (1997) address ways to represent actions and events in 'interval temporal logic,' building upon Allen's canonical work on the many ways to represent time (Allen, 1991). Their models also explore the complexities of discontinuous, overlapping, and nonlinear nature of time, with the goal of representation for computing purposes.

Concerns for time and temporality in the reuse of data are most often framed in terms of data 'life cycles.' The trio of short perspective articles that frame the scope of data science in the first issue of this journal illustrate contrasting perspectives even on this notion. Wing (2019), a computer scientist, took a linear view of data life cycles, emphasizing the privacy and ethical concerns at each stage. Leonelli (2019), a philosopher of science, explored a data-centric view of science, the ways in which data may be potential or actual evidence, and relationships between objects, data, models, knowledge, and interactions with the world. Borgman (2019), an information scientist, explored how data creation and stewardship vary by type of data and community, and over time, considering the 'lives and after lives' of data.

These are but a few of the many perspectives on data lifecycles, whose time stamps might refer to moment of production, of transmission, or reception, each of which may vary as communicated across devices, platforms, or networks, or as messages fragmented into bits and reassembled into packets in another time and place. Data archives and other records systems may stamp times of deposit, access, curation activities, duration of storage, and countless other transactions.

Leonelli and collaborators, in their work on 'data journeys,' are among the most explicit in studying temporal relationships in the creation and reuse of data. In their simplest formulation, the distance between data creator and reuser increases with time. When someone attempts to reuse data shortly after those resources become available, the reuser is most likely to have access to similar expertise and infrastructure. The greater the time elapsed between creation and reuse, the less likely that the data creation environment can be replicated and the more that the circumstances of data creation have changed (Leonelli, 2013, 2018; Leonelli & Tempini, 2020; Wylie, 2024).

Provenance is a category of metadata used to document relationships over time (Lynch, 2001; Mayernik, 2020a; Pasquier et al., 2017), and is a more elusive term than it may appear. The ability to trace temporal relationships such as origins, custody, amendments, ownership, transformations, or workflows, is essential to establishing chains of evidence and trust (Bettivia et al., 2022; Mayernik, 2019). As datasets are created, accessed, reused, modified, merged, split, and released for further reuse, the data journey may become an infinite regress.

Provenance metadata are computable representations of how a document or dataset came to be. Data provenance mechanisms such as the World Wide Web Consortium (W3C) recommendation on provenance, PROV (Moreau & Groth, 2013), use relative orderings of instantaneous events. The PROV method is based on Lamport's (1978) notion of logical clocks to avoid synchronization issues in distributed systems.

Version-controlled data, in which revisions or updates are captured as snapshots of a dataset at a particular moment, is another way to track provenance. Snapshots are ordered in relative, rather than absolute, time. Many version-control systems calculate or store differences between versions, accompanied with comments on how they vary. Version control may capture the existence of multiple instances of a dataset, but not capture, at least in a structured fashion, the processes that lead to the changes between datasets. While such temporal mechanisms are useful in data stewardship, they may not provide sufficient granularity for reusers to determine how, why, or whether multiple versions exist. An important example occurs in digital object identifiers (DOI) in which multiple versions of an object (such as a dataset or journal article) may be subsumed under a single DOI that references only the most recent version (DOI Foundation, 2023).

Temporal relationships within and between datasets are difficult to document or to represent (Drucker, 2009). The examples above are but a few of the ways in which data provenance records, as a form of metadata, can aid prospective reusers in situating data, identifying contexts in which data were collected, and thus bridge some of the temporal gaps between creators and reusers. When reusers obtain datasets that are recent in origin, conventions may be familiar,. Years later, it may become difficult to determine whether time stamps refer to Coordinated Universal Time, local time zones, or to assess the degree of accuracy in relative or absolute time. For example, while climate modelers are meticulous in their analyses, the longer and more geographically diverse the datasets on which they rely, the greater the degree of uncertainty they must address (Edwards, 2010).

In sum, temporal factors influence the distance between data creators and reusers in at least two ways. In one sense, the shorter the elapsed time between creation and reuse of dataset, the shorter the temporal distance for data exchange. In another sense, the more clear and precise the provenance information about a dataset, the shorter the distance may be between data creators and reusers.

## 3 Discussion

While data sharing requirements have become standard policy in most scholarly research domains, compliance varies widely by community, types of data, local infrastructure, enforcement, and many other factors. Not all data can or should be shared, nor can all data be sustained indefinitely. The costs of labor, technology, and infrastructure to share data are high, hard to measure, and unevenly distributed across stakeholders in the research enterprise.

Data creators make sharing choices in an environment of mixed incentives and multiple constraints. Promoting incentives to share data has been a challenge since the earliest days of open data policies. The burden of data sharing disproportionately falls on principal investigators, who may view the labor and costs of data release as a decrement to their research resources. Many are concerned about the ethics of inappropriate, misleading, or occasionally malicious reuse of their data (Borgman & Bourne, 2022; Hrynaszkiewicz et al., 2021; Martone et al., 2022; Staunton et al., 2021). Thus, enthusiasm for data sharing varies considerably by context, particularly by requirements of funding agencies and journals to which a researcher is subject, and by generation. Researchers who begin their careers in an era when they are expected to share data, and when they expect to have access to data for reuse, appear more amenable to sharing their data than those whose careers began in earlier decades under different norms.

Given that sharing research data is necessary, but not sufficient, for reuse, we argue for stronger research and policy emphases on data reuse. We ask who might reuse data, how, for what purposes, why, and when? Data sharing and reuse should not be viewed simply as a transactional relationship in which creators deposit and reusers retrieve and redeploy. Rather, open data policies should focus on knowledge exchange between those who create, reuse, and add value to research datasets.

These forms of knowledge exchange are best understood in the context of knowledge infrastructures that support data creation, sharing, reuse, and stewardship. Each stakeholder, whether data creator, reuser, archivist, funder, or any of the other actors involved in research data handling, relies upon layers of interconnected resources. These infrastructure components are human, technical, institutional, and often invisible. By surfacing these many interacting components, the dimensions of distance between data creators and reusers become more visible.

Distances between data creators and reusers are fundamentally social, embedded deeply in the methods and practices of creating knowledge within a research domain. These distance dimensions also are socio-technical in nature, because technologies can shorten distances by lubricating knowledge exchange and lengthen distances by causing friction between infrastructure components, including between people. In our theoretical framework, the shortest distance exists between data creators and reusers occurs when both have common expertise in a domain, are applying the same research methods, are collaborating on a project, curating their data to the highest standards of trust and stewardship, using data for the same purposes, and reusing data contemporaneously. As any of these six constraints are relaxed, the distance increases.

By addressing relationships between data creation and reuse, stakeholders in the research enterprise can invest more effectively in their data practices. While the distance construct suggests metrification of costs and positions of actors, we can offer only comparisons along a continuum. Attempts to identify specific costs of data production and reuse, such as the extensive study conducted by the U.S. National Academies on behalf of the National Institutes of Health, have focused on cost factors such as staffing expertise necessary for particular activities, and the salary rates of such staff (National Academies of Sciences, Engineering, and Medicine, 2020; Uffen & Kinkel, 2019) .

Although further investigation of data reuse cases is much needed, such studies are difficult to conduct even at small scales (Borgman et al., 2019). What we do know is that data reuse often is highly individualized. Researchers identify particular needs for data for specific purposes (Faniel & Jacobsen, 2010; Mayernik et al., 2008; Pasquetto et al., 2017, 2019; Wallis et al., 2010). Data practices within individual research teams, domains, institutions, and countries vary widely, as do knowledge infrastructures available to support those activities. We offer our theoretical model of the six dimensions of distance between data creators and reusers as a tool to think about these relationships. Our findings complement the FAIR principles that argue for making research data findable, accessible, interoperable, and reusable by others (Wilkinson et al., 2016). The FAIR principles are aspirational and socio-technical. They promote certain ways of conducting and disseminating research and they suggest architectures and software that can be deployed in support of these principles. Most arguments for the FAIR principles and related data sharing policies are based on transparency, reproducibility, equity, trust, and similar social concerns. Some open science arguments are economic, on the grounds that better access to data will promote reuse of existing resources and avoid duplication of effort. Others are based on commons arguments, that research products should be treated as public goods, which involves concerns for free riders and governance models.

Some shared data will be of high value to a diverse array of reusers for long periods of time. Other shared data will never be reused by anyone, ever. Core to the problem of determining what data to share and how to do so are the difficulties of evaluating what data are reused, at what rates, by whom, when, how, for what purposes, or to what effects. For those datasets in archives, downloads can be counted, but views and downloads are not equivalent to reuse. Someone may download a dataset and reuse it immediately, later, or never. Despite data citation now being technically feasible, with the advent of digital object identifiers (DOIs) assigned to datasets, adoption of data citation mechanisms by authors is minimal. Rarely do scholarly publications include references to datasets. Authors may reuse datasets or aggregated data without citing the data archive, individual datasets, or data creators (Borgman, 2016; Gregory et al., 2023).

## 4 Conclusions and Recommendations

We conclude first with a summary of our findings. Following the maxim that 'there is nothing so practical as a good theory[2],' (Lewin, 1951, p. 169), we conclude secondly with practical recommendations to stakeholders for how to employ our theory in data sharing and reuse.

### *4.1 Why Distance Matters*

Data creation, sharing, and reuse are inherently social processes. All rely on knowledge infrastructures, which in turn are complex and evolving. Many stakeholders are involved in these processes: not only the data creators and reusers, but archives, universities, funding agencies, publishers, policy makers, and private entities involved in research enterprise. The human infrastructure involved in data sharing and the technologies involved in making data FAIR (Wilkinson et al., 2016) have received the most attention, both in terms of policy and research.

Research on how, when, and why to reuse data deserves far more research attention. Never will it be possible to share all data, all the time, for all possible purposes, with all possible audiences. Hard choices must be made. Some data have high reuse value and are worthy of high investment. Other data may be of little value beyond the initial project in which they were created and thus worthy of minimal investment.

By developing the construct of distance between data creator and data reuser, our aim is to provoke new research questions, new research, and new investment, in effective and efficient circulation of research data. These dimensions form a theoretical framework to promote further research in ways to improve knowledge and data exchange, and to identify criteria for investments at each stage of data and research lifecycles.

Our first finding is that shared expertise in a research domain shortens distances between data creator and prospective reuser, thus increasing likelihood of knowledge exchange. By focusing on reuse within domains and communities, data creators may accelerate the reuse, and thus the research lifecycle, associated with the data they share.

Secondly, we find that data are most likely to be reused by those familiar with the research methods by which a dataset was created. By documenting their methods well, data creators can make their datasets reusable by a broader array of researchers.

Thirdly, we find that datasets often circulate most fluidly when collaborators have incentives to share data and to explain research contexts to each other. Thus researchers can employ their relationships with colleagues, students, and other collaborators to improve the flow of datasets and thus shorten distances between data creators and data reusers.

Fourth, we find that curation is essential to identifying, locating, retrieving, opening, processing, and reusing datasets. Data creators have the most intimate familiarity with their datasets and are in the best position to provide basic descriptive information about the origins, technologies, and processes necessary to reuse those data. Conversely, a lack of curation, especially the lack of useful metadata, can make datasets less discoverable and less reusable, thus increasing those distances. Archivists and other stakeholders in the lifecycle of a particular dataset may provide further curation, over the short and long term. Data creators influence the reusability of their datasets by their own investments in data curation and in their choices of data archiving. Greater curation investments shorten distances between the creators and prospective reusers.

---

[2] Frequently quoted by Everett M. Rogers in doctoral seminars at Stanford University. Rogers' (Rogers, 1995) theory of the diffusion of innovations influenced a vast array of scholarly domains.

Fifth, reusing datasets for purposes similar to those for which they were created is usually much easier than redeploying them for new purposes. Investments in data curation targeted at particular audiences that creators wish to reach can shorten distances in domain, methods, and other dimensions. Conversely, the lack of curation, especially the lack of useful metadata, can make datasets less discoverable and less reusable, thus increasing those distances and limiting the ability to expand the array of purposes for which a dataset might be reused.

Sixth and lastly, the influence of temporal factors on knowledge exchange is poorly understood but appears to be great. We identify two categories of temporal distance that influence distances between data creators and reusers. The first is the effect of elapsed time between creation and reuse of dataset on the likelihood of knowledge exchange. With time lapses of weeks or months, creators and prospective reusers can interact directly to exchange knowledge about a dataset. Knowledge infrastructures, including technologies and practices, increase in fragility over longer time frames. The second category is the role of curation, especially provenance information about a dataset, to shorten distances between data creators and reusers. Some datasets may remain useful for decades or centuries, in multiple domains, for multiple purposes, if sufficiently well curated.

Our claims about each of these distance dimensions can be treated as testable hypotheses, and we encourage others to test them in as many and varied contexts as possible. We call for further research on data reuse, and on recasting open science policies in terms of opportunities for transparency and reuse, rather than a narrow focus on data sharing. By focusing on knowledge exchange between data creators and reusers, and situating data reuse practices in the context of knowledge infrastructures, we expect research data practices to become more effective, efficient, and perhaps more cost-effective.

### *4.2 Recommendations to Stakeholders*

The six dimensions of distance between data creators and reusers can be employed as tools for self-reflection by stakeholders in the research enterprise. By understanding more about these dimensions, data creators can make better choices about how to manage, release, and share their data. Similarly, knowledge about distance dimensions may aid data reusers in finding, interpreting, and reusing available data – and in understanding the barriers to reusing existing data. Other stakeholders such as data archives, libraries, universities, funding agencies, and publishers, also can employ these distance dimensions in planning, priorities, and investments.

The questions to stakeholders in Table 1 summarize the specific recommendations below. These questions highlight expertise, knowledge, need for learning, and the relationship between creators and reusers, These questions foreground the 'invisible work' performed by data creators, data reusers, and many other individuals throughout the life cycles of research data (M. J. Bates, 1999; Borgman, 2003; Crain et al., 2016). For the sake of simplicity, our recommendations are grouped by four categories of stakeholders: data creators, data reusers, data archivists, and funding agencies. As explored throughout this article and in prior writing, we recognize that many stakeholders, and many relationships, are involved in knowledge exchanges. We emphasize the idea that **'it takes a village to share data'** (Borgman & Bourne, 2022; Borgman & Brand, 2022).

**Table 1 about here **

Table 1: Self-reflection questions for stakeholders. KI = knowledge infrastructure

| Self-Reflection Questions for Stakeholders by Distance Dimension | | | | |
|---|---|---|---|---|
| Dimension | Data Creator | Data Reuser | Data Archivist | Funding Agency |
| Domain | What is the scope of research domain in which you expect these data to be reusable? | What is the scope of research domain in which you expect to find reusable data? | What knowledge exchange practices are common in your domain? How is your data archive embedded in the KI of the domain? | What communities have effective KI to share and reuse data? To what extent do you expect data produced by a project to be reused inside or outside a domain? |
| Methods | How much knowledge of your methods is needed to interpret and reuse your data? How standard or innovative do you consider your methods to be? | For what methods do you have expertise and infrastructure to reuse data? How much time, labor, and resources are you willing to invest in data of unfamiliar methods? | How consistent are the methods applied in your domain? What array of methods knowledge exists in the communities of data creators and reusers you serve? | How do standard methods promote or constrain innovation in data reuse? What kinds of KI investments are needed to support reuse of data produced by disparate methods? |
| Collaboration | How well can your collaborators and students reuse your data? What KI do present or future collaborators need to interpret and reuse your data? | What do you need to know about your collaborators' data to reuse them? What common KI do you have to facilitate collaborative data reuse? | What information about a collaboration may aid in making these data more reusable? Can datasets be linked to communities of practice that might deposit or reuse these data? | How can collaborators advance their research through data exchange? How can agencies invest in KI to expand collaborations around reusable data? |
| Curation | What kinds of curation will add value to your data for reuse? What data archives are available to curate your data for reuse by your intended audiences? | What data archives might contain data of value for reuse? What ontologies may aid you in searching for reusable data? | What datasets are most worthy of curation investments for reuse? What knowledge do your reusers need to find, retrieve, and interpret archival data? | What KI investments in data curation will make data resources most reusable for domains or communities? |
| Purposes | For what array of purposes might your data be reusable by others? How can you curate your data to reach those audiences? | Where might data exist that were created for your intended purposes? In what other domains might data archives yield data reusable for your purposes? | How do the creation and reuse purposes of your contributors and searchers align? What curation might aid in knowledge exchange between them? | What KI investments would broaden the array of purposes for which data could be discovered and reused? |
| Time and Temporality | How long might your data be reusable? How much temporal precision is necessary to interpret and reuse your data? | For what time frames are data of interest for reuse? What KI do you have to interpret and reuse older data? | How long do you intend the data you curate to be reusable? What investments in temporal information might extend the life of your datasets? | What temporal factors are most relevant to data sharing and reuse? What KI investments will extend the reusability of research data? |

### *4.2.1 Recommendations to Data Creators*

Given that data creators cannot share all of their data, nor make them reusable by all people or machines, nor make those data useful for all time, we offer recommendations to guide creators in selecting what data to share and how to do so.

Researchers' concerns for how to share their data vary considerably by domain. For those in domains where data sharing and reuse are common, such as genomics and astronomy, our recommendations are to **identify the audience – the potential reusers – of your data as clearly, and as early in the research process, as possible**. A simple starting point is to consider prospective reusers at shortest distance from your work and the project at hand. Think of this audience as potential collaborators, and consider what knowledge of the research domain and of your research methods is necessary to reuse your data or to replicate your work. If your research practices build upon commonly available infrastructure such as instrumentation, software, and data archives; if your methods are relatively standard for the field; and your collaborations employ common data formats, then documenting your data for these audiences may be fairly straightforward.

Researchers who are net data creators, and whose data are frequently reused by others, such as those conducting sky surveys in astronomy or general social surveys in public policy, are more likely to have established knowledge infrastructures to support data release than those who share data only occasionally. If data creators wish to reach audiences in other domains, who rely on different infrastructures, use different methods and tools, or different data formats, then they should consider employing interface tools that can bridge gaps between data formats and other technical methods that promote portability across infrastructures. The more innovative a research team's methods, tools, software, and instrumentation, the more documentation for reuse that may be required to reach new audiences. Similarly, to make data reusable for longer time frames, more provenance information is necessary. Hardware, software, instrumentation, and tools associated with digital data do not age well. Datasets that are migrated regularly to current environments will remain useful for longer; resurrecting datasets left untouched for a few years becomes increasingly difficult. Data creators should bear in mind how much audiences change over time. Today's data may cease to be useful, whereas old data may become new again, for very different uses and users.

Lastly, we recommend that **data creators find ways to test the reusability of datasets that they share.** By asking students or collaborators to find, open, and reuse a dataset, data creators can learn how well they have documented and formatted their data resources. Reproducing a study based on a journal article and associated public dataset can be an effective pedagogical method to teach students about research methods and data management. Similarly, mutual testing of the reusability of datasets can enhance knowledge exchange among collaborators.

### *4.2.2 Recommendations to Data Reusers*

Our recommendations to prospective data reusers **are to consider where you reside on each of the six dimensions of distance.** In some areas of scholarship, data reuse is sufficiently common that many researchers are net data reusers. Data archives that support space observatories with long missions, such as the Hubble Space Telescope, now yield more journal articles than do new data collection. Many areas of biosciences rely heavily on public data archives of gene sequences, clinical trials, or similar resources.

Finding data within the same domain is generally easiest, given a prospective data reuser's likely knowledge of methods, terminology, technology, and data sources. Obtaining data from current collaborators, or establishing new collaborations for the purposes of sharing data, is another way to shorten distances from origin. In domains such as astronomy and genomics, data sharing is normal practice. These are domains with extensive knowledge infrastructures for creating, sharing, and reusing data with standardized data formats and tools. However, even in astronomy and genomics, making sense of data produced a decade earlier can be difficult.

Given that data reuse is a form of knowledge transfer from the data creators, any act of data reuse may involve considerable learning about the origins of the data. Learning may involve reading literature, obtaining software and instrumentation, hiring staff with appropriate skills, or contacting the data creators who have the most intimate knowledge of their origins. Hence, data reusers should budget for this learning process.

#### *4.2.3 Recommendations to Data Archivists*

Of the many stakeholders in data creation and reuse, archivists usually are the most aware of the distances, and most attentive to metadata. They may be less aware of the levels of expertise along these dimensions of those who submit datasets or retrieve data from their archives.

The role of data archivists varies greatly by type of archive, as discussed in Section 2.4, *Curation Distance.* Archivists play the most active role in 'reference data collections' that are core resources for a domain. They work closely with data creators to curate datasets by augmenting metadata and documentation, and by packaging datasets for ingest, a process that may take days or months, depending upon the degree of interaction with submitters that is required. They also provide assistance in searching for datasets appropriate for a particular reuse. At the other extreme are self-curated repositories, which may employ archivists only to manage the processing, and who have minimal interaction with data creators or prospective reusers. Archivists play important instructional roles along the spectrum of data creation and reuse.

Our general recommendations to data archivists are both to **'know thy user' and to recognize how difficult it may be to know enough about your data contributors and data retrievers to bridge the distances**. Domain-specific archives, whether in astronomy or art history, are in the strongest position to construct coherent collections and to know their user communities. The curation services provided by domain archives are essential to knowledge transfer between data creators and reusers. Toward sustaining and improving curation services, these archives can **attend to documenting as much information as possible along each of these distance dimensions and their technical aspects**: infrastructure components of the domain on which data interpretation depends; methods and technical standards associated with a dataset; collaborative project or research center and associated data formats; documentation provided by contributors plus supplemental curation by the archive; purposes for which data were created and any known interfaces between tools or environments; and time frame of data creation with as much provenance information as can be obtained.

This recommendation for more documentation **argues for a careful selection procedure by archives**. Explicit tradeoffs may be necessary between acquiring greater volumes of data and investing in more extensive curation of fewer datasets.

While we are recommending more metadata than is usually provided for individual datasets, and recognizing that metadata creation is expensive and time-consuming, these recommendations also can be applied to guidance for self-curation by data contributors. Generic data archives such as institutional repositories rely almost entirely on self-curation by data creators who contribute datasets for longer term storage and stewardship. These data archives may not be able to invest in curation of individual datasets, **but they can offer instruction and guidance in how best to curate datasets to improve the abilities of others to reuse them**.

#### *4.2.4 Recommendations to Research Funding Agencies*

Our recommendations to research funding agencies are manifold. Funding agency policies focus primarily on data sharing, placing responsibility for data release on the principal investigators of grants. While releasing data is prerequisite to reuse, we recommend greater emphasis on data reuse, and on more wholistic approaches to building the infrastructures and relationships between stakeholders necessary to bridge the distances between data creators and prospective data reusers. A good start is to **recognize that 'it takes a village to share data'** and that better partnerships within and between universities are needed (Borgman & Bourne, 2022; Borgman & Brand, 2022). Additionally, funding calls could ask principal investigators to **identify intended reusers of their data** and the corresponding investments needed to bridge these distances.

We recommend **further support to address the gap in knowledge about which activities lead to data reuse.** This is in addition to the roles and work necessary to facilitate data reuse. Here, further empirical evidence and rich qualitative studies are warranted.

Funding agencies in the U.S. and Europe, which are the communities we know best, are investing in research methods, in collaborations, and in each of the technical areas necessary to bridge these distances, albeit more within than between domains. We **recommend a broader focus on the human infrastructure necessary to support the knowledge infrastructures necessary for effective data exchange**.

## 5 Acknowledgements

We kindly acknowledge the contributions of colleagues who read drafts, discussed sections, and commented substantively on public presentations of this article, including Karen S. Baker, Michael K. Buckland, Johanna Drucker, Andrea M. Ghez, Kathleen M. Gregory, James Howison, Clifford A. Lynch, Matthew S. Mayernik, Xiao-Li Meng, Andrea Scharnhorst, and three anonymous reviewers for the journal.

## 6 References

Acker, A. (2015). Toward a Hermeneutics of Data. *IEEE Annals of the History of Computing*, *37*(3), 70–75. https://doi.org/10.1109/MAHC.2015.68

Acker, A. (2017). When is a record? A Research Framework for Locating Electronic Records in Infrastructure. In A. J. Gilliland, S. McKemmish, & A. J. Lau (Eds.), *Research in the Archival Multiverse* (pp. 288–323). Monash University Press.

Alipourfard, N., Arendt, B., Benjamin, D. J., Benkler, N., Bishop, M., Burstein, M., Bush, M., Caverlee, J., Chen, Y., Clark, C., Almenberg, A. D., Errington, T., Fidler, F., Fox [SCORE, N., Frank, A., Fraser, H., Friedman, S., Gelman, B., Gentile, J., … Wu, J. (2021). *Systematizing Confidence in Open Research and Evidence (SCORE)*. SocArXiv. https://doi.org/10.31235/osf.io/46mnb

Allen, J. F. (1991). Time and time again: The many ways to represent time. *International Journal of Intelligent Systems*, *6*(4), 341–355. https://doi.org/10.1002/int.4550060403

Allen, J. F., & Ferguson, G. (1997). Actions and Events in Interval Temporal Logic. In O. Stock (Ed.), *Spatial and Temporal Reasoning* (pp. 205–245). Springer Netherlands. https://doi.org/10.1007/978-0-585-28322-7_7

Asok, K., Dandpat, S. S., Gupta, D. K., & Shrivastava, P. (2024). Common Metadata Framework for Research Data Repository: Necessity to Support Open Science. *Journal of Library Metadata*, *24*(2), 133–145. https://doi.org/10.1080/19386389.2024.2329370

Aspesi, C., & Brand, A. (2020). In pursuit of open science, open access is not enough. *Science*, *368*(6491), 574–577. https://doi.org/10.1126/science.aba3763

Baack, S. (2024). *Training Data for the Price of a Sandwich: Common Crawl's Impact on Generative AI*. https://foundation.mozilla.org/en/research/library/generative-ai-training-data/common-crawl/

Baker, K. S., Duerr, R. E., & Parsons, M. A. (2016). Scientific Knowledge Mobilization: Co-evolution of Data Products and Designated Communities. *International Journal of Digital Curation*, *10*(2), 110–135. https://doi.org/10.2218/ijdc.v10i2.346

Baker, K. S., & Mayernik, M. S. (2020). Disentangling knowledge production and data production. *Ecosphere*, *11*(7), e03191. https://doi.org/10.1002/ecs2.3191

Baker, K. S., & Yarmey, L. (2009). Data Stewardship: Environmental Data Curation and a Web-of-Repositories. *International Journal of Digital Curation*, *4*(2), 12–27. https://doi.org/10.2218/ijdc.v4i2.90

Barker, M., Chue Hong, N. P., Katz, D. S., Lamprecht, A.-L., Martinez-Ortiz, C., Psomopoulos, F., Harrow, J., Castro, L. J., Gruenpeter, M., Martinez, P. A., & Honeyman, T. (2022). Introducing the FAIR Principles for research software. *Scientific Data*, *9*(1), Article 1. https://doi.org/10.1038/s41597-022-01710-x

Bastow, R., & Leonelli, S. (2010). Sustainable digital infrastructure. *EMBO Reports*, *11*(10), 730–734. https://doi.org/10.1038/embor.2010.145

Bates, J., Lin, Y.-W., & Goodale, P. (2016). Data journeys: Capturing the socio-material constitution of data objects and flows. *Big Data & Society*, *3*(2), 2053951716654502. https://doi.org/10.1177/2053951716654502

Bates, M. J. (1999). The invisible substrate of information science. *Journal of the American Society for Information Science*, *50*(12), 1043–1050. https://doi.org/10.1002/(SICI)1097-4571(1999)50:12<1043::AID-ASI1>3.0.CO;2-X

Benjelloun, O., Chen, S., & Noy, N. (2020). Google Dataset Search by the Numbers. In J. Z. Pan, V. Tamma, C. d'Amato, K. Janowicz, B. Fu, A. Polleres, O. Seneviratne, & L. Kagal (Eds.), *The Semantic Web – ISWC 2020* (Vol. 12507, pp. 667–682). Springer International Publishing. https://doi.org/10.1007/978-3-030-62466-8_41

Bettivia, R., Cheng, Y.-Y., & Gryk, M. R. (2022). *Documenting the Future: Navigating Provenance Metadata Standards*. Springer Nature.

Bierer, B. E., Crosas, M., & Pierce, H. H. (2017). Data Authorship as an Incentive to Data Sharing. *New England Journal of Medicine*, *376*. https://doi.org/10.1056/NEJMsb1616595

Blair, A. M. (2010). *Too Much to Know: Managing Scholarly Information before the Modern Age*. Yale University Press.

Bly, A., & Brand, A. (2023). AI is muddying the truth. We've known how to fix it for centuries. *BostonGlobe.Com*. https://www.bostonglobe.com/2023/08/07/magazine/ai-is-muddying-truth-weve-known-how-fix-it-centuries/

Borgman, C. L. (2000). *From Gutenberg to the Global Information Infrastructure: Access to Information in the Networked World*. MIT Press.

Borgman, C. L. (2003). The Invisible Library: Paradox of the Global Information Infrastructure. *Library Trends*, *51*, 652–674. http://hdl.handle.net/2142/8487

Borgman, C. L. (2007). *Scholarship in the Digital Age: Information, Infrastructure, and the Internet*. MIT Press.

Borgman, C. L. (2015). *Big Data, Little Data, No Data: Scholarship in the Networked World*. MIT Press.

Borgman, C. L. (2016). *Data citation as a bibliometric oxymoron*. https://escholarship.org/uc/item/8w36p9zf

Borgman, C. L. (2019). The lives and after lives of data. *Harvard Data Science Review*, *1*(1). https://doi.org/10.1162/99608f92.9a36bdb6

Borgman, C. L., & Bourne, P. E. (2022). Why It Takes a Village to Manage and Share Data. *Harvard Data Science Review*, *4*(3). https://doi.org/10.1162/99608f92.42eec111

Borgman, C. L., & Brand, A. (2022). Data blind: Universities lag in capturing and exploiting data. *Science*, *378*(6626), 1278–1281. https://doi.org/10.1126/science.add2734

Borgman, C. L., & Brand, A. (2024). The Future of Data in Research Publishing: From Nice to Have to Need to Have? *Harvard Data Science Review*, *Special Issue 4*. https://doi.org/10.1162/99608f92.b73aae77

Borgman, C. L., Darch, P. T., Sands, A. E., & Golshan, M. S. (2016). The durability and fragility of knowledge infrastructures: Lessons learned from astronomy. *Proceedings of the Association for Information Science and Technology*, *53*, 1–10. http://dx.doi.org/10.1002/pra2.2016.14505301057

Borgman, C. L., Golshan, M. S., Sands, A. E., Wallis, J. C., Cummings, R. L., Darch, P. T., & Randles, B. M. (2016). Data Management in the Long Tail: Science, Software, and Service. *International Journal of Digital Curation*, *11*(1), 128–149. https://doi.org/10.2218/ijdc.v11i1.428

Borgman, C. L., Scharnhorst, A., & Golshan, M. S. (2019). Digital data archives as knowledge infrastructures: Mediating data sharing and reuse. *Journal of the Association for Information Science and Technology*, *70*(8), 888–904. https://doi.org/10.1002/asi.24172

Borgman, C. L., & Wofford, M. F. (2021). From Data Processes to Data Products: Knowledge Infrastructures in Astronomy. *Harvard Data Science Review*, *3*(3). https://doi.org/10.1162/99608f92.4e792052

Bos, N. M., Zimmerman, A. S., Olson, J. S., Yew, J., Yerkie, J., Dahl, E., Cooney, D., & Olson, G. M. (2008). From Shared Databases to Communities of Practice: A Taxonomy of Collaboratories. In G. M. Olson, A. S. Zimmerman, & N. M. Bos (Eds.), *Scientific Collaboration on the Internet* (pp. 52–72). MIT Press. https://doi.org/10.7551/mitpress/9780262151207.003.0004

Bourne, P. E. (2005). Will a biological database be different from a biological journal? *PLoS Computational Biology*, *1*. http://dx.doi.org/10.1371/journal.pcbi.0010034

Bowers, S., McPhillips, T., Ludäscher, B., Cohen, S., & Davidson, S. B. (2006). A Model for User-Oriented Data Provenance in Pipelined Scientific Workflows. In L. Moreau & I. Foster (Eds.), *Provenance and Annotation of Data* (pp. 133–147). Springer. https://doi.org/10.1007/11890850_15

Bowker, G. C. (2005). *Memory Practices in the Sciences*. MIT Press.

Brand, A. (2022, April 8). *Open access loses when publishers are vilified*. Times Higher Education (THE). https://www.timeshighereducation.com/opinion/open-access-loses-when-publishers-are-vilified

Brickley, D., Burgess, M., & Noy, N. (2019). Google Dataset Search: Building a search engine for datasets in an open Web ecosystem. *The World Wide Web Conference*, 1365–1375. https://doi.org/10.1145/3308558.3313685

Brinckman, A., Chard, K., Gaffney, N., Hategan, M., Jones, M. B., Kowalik, K., Kulasekaran, S., Ludäscher, B., Mecum, B. D., Nabrzyski, J., Stodden, V., Taylor, I. J., Turk, M. J., & Turner, K. (2019). Computing environments for reproducibility: Capturing the "Whole Tale." *Future Generation Computer Systems*, *94*, 854–867. https://doi.org/10.1016/j.future.2017.12.029

Castelo, S., Rampin, R., Santos, A., Bessa, A., Chirigati, F., & Freire, J. (2021). Auctus: A dataset search engine for data discovery and augmentation. *Proceedings of the VLDB Endowment*, *14*(12), 2791–2794. https://doi.org/10.14778/3476311.3476346

Castro, E., Crosas, M., Garnett, A., Sheridan, K., & Altman, M. (2017). Evaluating and Promoting Open Data Practices in Open Access Journals. *Journal of Scholarly Publishing*, *49*(1), 66–88. https://doi.org/10.3138/jsp.49.1.66

Chapman, A., Simperl, E., Koesten, L., Konstantinidis, G., Ibáñez, L.-D., Kacprzak, E., & Groth, P. (2020). Dataset search: A survey. *The VLDB Journal*, *29*(1), 251–272. https://doi.org/10.1007/s00778-019-00564-x

Chen, X., Dallmeier-Tiessen, S., Dasler, R., Feger, S., Fokianos, P., Gonzalez, J. B., Hirvonsalo, H., Kousidis, D., Lavasa, A., Mele, S., Rodriguez, D. R., Šimko, T., Smith, T., Trisovic, A., Trzcinska, A., Tsanaktsidis, I., Zimmermann, M., Cranmer, K., Heinrich, L., … Neubert, S. (2018). Open is not enough. *Nature Physics*, 1. https://doi.org/10.1038/s41567-018-0342-2

Choi, A., & Xin, X. (2021). Data Curation in Practice: Extract Tabular Data from PDF Files Using a Data Analytics Tool. *Journal of eScience Librarianship*, *10*(3). https://doi.org/10.7191/jeslib.2021.1209

Christen, P., & Schnell, R. (2024). When Data Science Goes Wrong: How Misconceptions About Data Capture and Processing Causes Wrong Conclusions. *Harvard Data Science Review*, *6*(1). https://doi.org/10.1162/99608f92.34f8e75b

Clement, T., & Acker, A. (2019). Data Cultures, Culture as Data – Special Issue of Cultural Analytics. *Journal of Cultural Analytics*. https://doi.org/10.22148/16.035

Collins, H. M., & Evans, R. (2002). The Third Wave of Science Studies: Studies of Expertise and Experience. *Social Studies of Science*, *32*(2), 235–296. http://sss.sagepub.com/content/32/2/235.short

Collins, H. M., & Evans, R. (2007). *Rethinking Expertise*. University of Chicago Press.

Collins, H. M., Evans, R., & Gorman, M. E. (2007). Trading zones and interactional expertise. *Studies in History and Philosophy of Science Part A*, *38*(4), 657–666. https://doi.org/10.1016/j.shpsa.2007.09.003

*Common Crawl*. (2023). Data Resource. https://commoncrawl.org/

Consultative Committee for Space Data Systems. (2012). *Reference model for an Open Archival Information System (OAIS)* (Recommendation for Space Data System Practices CCSDS 650.0-M-2 Magenta Book; Recommended Practice, Issue 2). https://public.ccsds.org/pubs/650x0m2.pdf

*CoreTrustSeal*. (2024). https://www.coretrustseal.org/

Crain, M., Poster, W., & Cherry, M. (2016). *Invisible Labor: Hidden Work in the Contemporary World*. University of California Press.

Crane, D. (1972). *Invisible Colleges: Diffusion of Knowledge in Scientific Communities*. University of Chicago Press.

Darch, P. T. (2016, March 10). Many Methods, Many Microbes: Methodological Diversity and Standardization in the Deep Subseafloor Biosphere. *iConference 2016 Proceedings*. iConference 2016: Partnership with Society, Philadelphia, USA. https://doi.org/10.9776/16246

Darch, P. T., & Borgman, C. L. (2016). Ship space to database: Emerging infrastructures for studies of the deep subseafloor biosphere. *PeerJ Computer Science*, *2*, e97. https://doi.org/10.7717/peerj-cs.97

Darch, P. T., Borgman, C. L., Traweek, S., Cummings, R. L., Wallis, J. C., & Sands, A. E. (2015). What lies beneath?: Knowledge infrastructures in the subseafloor biosphere and beyond. *International Journal on Digital Libraries*, *16*(1), 61–77. https://doi.org/10.1007/s00799-015-0137-3

Darch, P. T., Sands, A. E., Borgman, C. L., & Golshan, M. S. (2020). Library cultures of data curation: Adventures in astronomy. *Journal of the Association for Information Science and Technology*, *71*(12), 1470–1483. https://doi.org/10.1002/asi.24345

Darch, P. T., Sands, A. E., Borgman, C. L., & Golshan, M. S. (2021). Do the stars align?: Stakeholders and strategies in libraries' curation of an astronomy dataset. *Journal of the Association for Information Science and Technology*, *72*(2), 239–252. https://doi.org/10.1002/asi.24392

Daston, L. (Ed.). (2017). *Science in the Archives: Pasts, Presents, Futures*. University of Chicago Press.

Daston, L. (2023). *Rivals: How Scientists Learned to Cooperate*. Columbia Global Reports.

Data Archiving and Networked Services. (2024). *DANS: Centre of expertise & repository for research data*. DANS. https://dans.knaw.nl/en/

David, R., Rybina, A., Burel, J.-M., Heriche, J.-K., Audergon, P., Boiten, J.-W., Coppens, F., Crockett, S., Exter, K., Fahrner, S., Fratelli, M., Goble, C., Gormanns, P., Grantner, T., Grüning, B., Gurwitz, K. T., Hancock, J. M., Harmse, H., Holub, P., … Gribbon, P. (2023). "Be sustainable": EOSC-Life recommendations for implementation of FAIR principles in life science data handling. *The EMBO Journal*, *42*(23), e115008. https://doi.org/10.15252/embj.2023115008

Davidson, E., Wessel, L., Winter, J. S., & Winter, S. (2023). Future directions for scholarship on data governance, digital innovation, and grand challenges. *Information and Organization*, *33*(1), 100454. https://doi.org/10.1016/j.infoandorg.2023.100454

Directorate-General for Research and Innovation, European Commission, EOSC Executive Board, Corcho, O., Eriksson, M., Kurowski, K., Ojsteršek, M., Choirat, C., Sanden, M. van de, & Coppens, F. (2021). *EOSC interoperability framework: Report from the EOSC Executive Board Working Groups FAIR and Architecture*. Publications Office of the European Union. https://data.europa.eu/doi/10.2777/620649

Distance. (2024). In *Oxford English Dictionary*. Oxford University Press. https://www.oed.com/search/dictionary/?q=distance

DOI Foundation. (2023). *DOI (Digital Object Identifier) Handbook*. https://doi.org/10.1000/182

Donaldson, D. R., & Koepke, J. W. (2022). A focus groups study on data sharing and research data management. *Scientific Data*, *9*(1), 345. https://doi.org/10.1038/s41597-022-01428-w

Drucker, J. (2009). Temporal Modeling. In J. Drucker (Ed.), *SpecLab: Digital Aesthetics and Projects in Speculative Computing* (p. 0). University of Chicago Press. https://doi.org/10.7208/chicago/9780226165097.003.0003

Edwards, P. N. (2010). *A vast machine: Computer models, climate data, and the politics of global warming*. MIT Press.

Edwards, P. N., Mayernik, M. S., Batcheller, A. L., Bowker, G. C., & Borgman, C. L. (2011). Science Friction: Data, Metadata, and Collaboration. *Social Studies of Science*, *41*(5), 667–690. https://doi.org/10.1177/0306312711413314

Eschenfelder, K. R., Shankar, K., & Downey, G. (2022). The financial maintenance of social science data archives: Four case studies of long-term infrastructure work. *Journal of the Association for Information Science and Technology*, *73*(12), 1723–1740. https://doi.org/10.1002/asi.24691

Faniel, I. M., & Jacobsen, T. E. (2010). Reusing Scientific Data: How Earthquake Engineering Researchers Assess the Reusability of Colleagues' Data. *Journal of Computer Supported Cooperative Work*, *19*(3–4), 355–375. https://doi.org/10.1007/s10606-010-9117-8

Faniel, I. M., Kriesberg, A., & Yakel, E. (2016). Social scientists' satisfaction with data reuse. *Journal of the Association for Information Science and Technology*, *67*(6), 1404–1416. https://doi.org/10.1002/asi.23480

Faniel, I. M., & Yakel, E. (2017). Practices Do Not Make Perfect: Disciplinary Data Sharing and Reuse Practices and Their Implications for Repository Data Curation. In L. R. Johnston (Ed.), *Curating Research Data, Volume One: Practical Strategies for Your Digital Repository* (pp. 103–126). Association of College and Research Libraries. http://www.oclc.org/research/publications/2017/practices-do-not-make-perfect.html

*Figshare*. (2024). Digital Science. https://www.digital-science.com/product/figshare/

Frenken, K., Hoekman, J., Kok, S., Ponds, R., van Oort, F., & van Vliet, J. (2009). Death of Distance in Science? A Gravity Approach to Research Collaboration. In A. Pyka & A. Scharnhorst (Eds.), *Innovation Networks: New Approaches in Modelling and Analyzing* (pp. 43–57). Springer. https://doi.org/10.1007/978-3-540-92267-4_3

Galison, P. (1997). *Image and Logic: A Material Culture of Microphysics*. University Of Chicago Press.

Gebel, L., Velu, C., & Vidal-Puig, A. (2024). The strategy behind one of the most successful labs in the world. *Nature*, *630*(8018), 813–816. https://doi.org/10.1038/d41586-024-02085-2

Genova, F. (2013). Strasbourg Astronomical Data Center (CDS). *Data Science Journal*, *12*, WDS56–WDS60. https://doi.org/10.2481/dsj.WDS-007

Genova, F. (2018). Data as a Research Infrastructure—CDS, the Virtual Observatory, Astronomy, and beyond. *EPJ Web of Conferences*, *186*, 01001. https://doi.org/10.1051/epjconf/201818601001

Getty Research Institute. (2008). *Introduction to metadata* (M. Baca, Ed.; 2nd ed). Getty Research Institute. http://www.getty.edu/research/publications/electronic_publications/intrometadata/index.html

Gilliland-Swetland, A. J. (2000). *Enduring Paradigm, New Opportunities: The Value of the Archival Perspective in the Digital Environment.* Council on Library and Information Resources. http://www.clir.org/pubs/reports/pub89/contents.html

Ginsparg, P. (2021). Lessons from arXiv's 30 years of information sharing. *Nature Reviews Physics*, *3*(9), Article 9. https://doi.org/10.1038/s42254-021-00360-z

Gregory, K. M., Groth, P., Cousijn, H., Scharnhorst, A., & Wyatt, S. (2019). Searching Data: A Review of Observational Data Retrieval Practices in Selected Disciplines. *Journal of the Association for Information Science and Technology*, *70*(5), 419–432. https://doi.org/10.1002/asi.24165

Gregory, K. M., Groth, P., Scharnhorst, A., & Wyatt, S. (2020). Lost or Found? Discovering Data Needed for Research. *Harvard Data Science Review*, *2*(2). https://doi.org/10.1162/99608f92.e38165eb

Gregory, K. M., & Koesten, L. (2022). Introduction. In K. Gregory & L. Koesten (Eds.), *Human-Centered Data Discovery* (pp. 1–6). Springer International Publishing. https://doi.org/10.1007/978-3-031-18223-5_1

Gregory, K. M., Ninkov, A., Ripp, C., Roblin, E., Peters, I., & Haustein, S. (2023). Tracing data: A survey investigating disciplinary differences in data citation. *Quantitative Science Studies*, *4*(3), 622–649. https://doi.org/10.1162/qss_a_00264

Griesemer, J. R. (2020). A Data Journey Through Dataset-Centric Population Genomics. In S. Leonelli & N. Tempini (Eds.), *Data Journeys in the Sciences* (pp. 145–170). Springer International Publishing. https://doi.org/10.1007/978-3-030-37177-7_3

Groth, P., Cousijn, H., Clark, T., & Goble, C. A. (2019). FAIR Data Reuse – the Path through Data Citation. *Data Intelligence*, 78–86. https://doi.org/10.1162/dint_a_00030

Harper, L. M. (2023). *Data Reuse Among Digital Humanities Scholars: A Qualitative Study of Practices, Challenges and Opportunities*. https://doi.org/10.20381/RUOR-29651

Hilgartner, S., & Brandt-Rauf, S. I. (1994). Data Access, Ownership, and Control: Toward Empirical Studies of Access Practices. *Science Communication*, *15*(4), 355–372. https://doi.org/10.1177/107554709401500401

Howison, J., & Bullard, J. (2016). Software in the scientific literature: Problems with seeing, finding, and using software mentioned in the biology literature. *Journal of the Association for Information Science and Technology*, *67*(9), 2137–2155. https://doi.org/10.1002/asi.23538

Hrynaszkiewicz, I., Harney, J., & Cadwallader, L. (2021). A Survey of Researchers' Needs and Priorities for Data Sharing. *Data Science Journal*, *20*(1). https://doi.org/10.5334/dsj-2021-031

Hudson Vitale, C. (2023). *The Quality and Cost of Shared Data*. https://hdl.handle.net/11299/255741

Interuniversity Consortium for Political and Social Research. (2024). *ICPSR Data Excellence Research Impact*. https://www.icpsr.umich.edu/web/pages/

Jackson, S. J., Ribes, D., & Buyuktur, A. (2010). Exploring Collaborative Rhythm: Temporal Flow and Alignment in Collaborative Scientific Work. *iConference 2010 Proceedings*. iConference, Urbana-Champaign, IL. https://www.ideals.illinois.edu/handle/2142/14955

Jia, H., Miller, L. I., Hicks, J., Moscot, E., Landberg, A., Heflin, J., & Davison, B. D. (2022). Truth in a sea of data: Adoption and use of data search tools among researchers and journalists. *Information, Communication & Society*, *0*(0), 1–20. https://doi.org/10.1080/1369118X.2022.2147398

Jirotka, M., Lee, C. P., & Olson, G. M. (2013). Supporting Scientific Collaboration: Methods, Tools and Concepts. *Computer Supported Cooperative Work (CSCW)*, *22*(4–6), 667–715. https://doi.org/10.1007/s10606-012-9184-0

Johnston, L. R., Carlson, J., Hudson-Vitale, C., Imker, H., Kozlowski, W., Olendorf, R., Stewart, C., Blake, M., Herndon, J., McGeary, T. M., & Hull, E. (2018). Data Curation Network: A Cross-Institutional Staffing Model for Curating Research Data. *International Journal of Digital Curation*, *13*(1), Article 1. https://doi.org/10.2218/ijdc.v13i1.616

Johnston, L. R., Curty, R., Braxton, S. M., Carlson, J., Hadley, H., Lafferty-Hess, S., Luong, H., Petters, J. L., & Kozlowski, W. A. (2024a). Understanding the value of curation: A survey of US data repository curation practices and perceptions. *PLOS ONE*, *19*(6), e0301171. https://doi.org/10.1371/journal.pone.0301171

Johnston, L. R., Curty, R., Braxton, S. M., Carlson, J., Hadley, H., Lafferty-Hess, S., Luong, H., Petters, J. L., & Kozlowski, W. A. (2024b). Understanding the value of curation: A survey of US data repository curation practices and perceptions. *PLOS ONE*, *19*(6), e0301171. https://doi.org/10.1371/journal.pone.0301171

Jung, J. Y., Steinberger, T., King, J. L., & Ackerman, M. S. (2022). How Domain Experts Work with Data: Situating Data Science in the Practices and Settings of Craftwork. *Proceedings of the ACM on Human-Computer Interaction*, *6*(CSCW1), 58:1-58:29. https://doi.org/10.1145/3512905

Katz, D. S., Gruenpeter, M., & Honeyman, T. (2021). Taking a fresh look at FAIR for research software. *Patterns*, *2*(3). https://doi.org/10.1016/j.patter.2021.100222

Katz, D. S., & Hong, N. P. C. (2024). Special issue on software citation, indexing, and discoverability. *PeerJ Computer Science*, *10*, e1951. https://doi.org/10.7717/peerj-cs.1951

Knorr-Cetina, K. (1999). *Epistemic Cultures: How the Sciences Make Knowledge*. Harvard University Press.

Koesten, L., Gregory, K. M., Groth, P., & Simperl, E. (2021). Talking datasets – Understanding data sensemaking behaviours. *International Journal of Human-Computer Studies*, *146*, 102562. https://doi.org/10.1016/j.ijhcs.2020.102562

Koesten, L., Vougiouklis, P., Simperl, E., & Groth, P. (2020). Dataset Reuse: Toward Translating Principles to Practice. *Patterns*, 100136. https://doi.org/10.1016/j.patter.2020.100136

Kruithof, E. H., Vanroelen, C., & Borre, L. V. den. (2024). Five Suggestions Towards User-Centred Data Repositories in the Social Sciences. *Data Science Journal*, *23*(1). https://doi.org/10.5334/dsj-2024-019

Lagoze, C. (2001). The open archives initiative: Building a low-barrier interoperability framework. *Proceedings of the First ACM/IEEE-CS Joint Conference on Digital Libraries*, 24–28.

Lakoff, G., & Johnson, M. (2003). *Metaphors We Live By* (W. a new Afterword, Ed.). University of Chicago Press. https://press.uchicago.edu/ucp/books/book/chicago/M/bo3637992.html

Lampland, M., & Star, S. L. (2009). *Standards and Their Stories: How Quantifying, Classifying, and Formalizing Practices Shape Everyday Life*. Cornell University Press.

Lamport, L. (1978). Time, clocks, and the ordering of events in a distributed system. *Communications of the ACM*, *21*(7), 558–565. https://doi.org/10.1145/359545.359563

Lane, J., & Potok, N. (2024). Democratizing Data: Our Vision. *Harvard Data Science Review*, *Special Issue 4*. https://doi.org/10.1162/99608f92.106473cf

Laney, D. (2001). *3D Data Management: Controlling Data Volume, Velocity and Variety* (Application Delivery Strategies File 949). META Group (Gartner). http://blogs.gartner.com/doug-laney/files/2012/01/ad949-3D-Data-Management-Controlling-Data-Volume-Velocity-and-Variety.pdf

Latour, B. (1987). *Science in action: How to follow scientists and engineers through society*. Harvard University Press.

Latour, B. (1993). *We Have Never Been Modern* (C. Porter, Trans.). Harvard University Press.

Latour, & Woolgar, S. (1979). *Laboratory Life: The Social Construction of Scientific Facts*. Sage.

Lautenbacher, L., Samaras, P., Muller, J., Grafberger, A., Shraideh, M., Rank, J., Fuchs, S. T., Schmidt, T. K., The, M., Dallago, C., Wittges, H., Rost, B., Krcmar, H., Kuster, B., & Wilhelm, M. (2022). ProteomicsDB: Toward a FAIR open-source resource for life-science research. *Nucleic Acids Research*, *50*(D1), D1541–D1552. https://doi.org/10.1093/nar/gkab1026

Lave, J., & Wenger, E. (1991). *Situated Learning: Legitimate Peripheral Participation*. Cambridge University Press.

Lawrence, A. (2022). From Pamphlets to PDFs: The Shadow Histories of Research Publishing. *Pop! Public. Open. Participatory*, *4*. https://doi.org/10.54590/pop.2022.006.

Lee, S., & Dourish, P. (2024). Reconfiguring Data Relations: Institutional Dynamics around Data in Local Governance. *Proc. ACM Hum.-Comput. Interact.*, *8*(CSCW2), 420:1-420:28. https://doi.org/10.1145/3686959

Leibniz Institute for the Social Sciences. (2024). *GESIS*. https://www.gesis.org/en/home

Leonelli, S. (2013). Integrating data to acquire new knowledge: Three modes of integration in plant science. *Studies in History and Philosophy of Science Part C: Studies in History and Philosophy of Biological and Biomedical Sciences*, *44*(4), 503–514. https://doi.org/10.1016/j.shpsc.2013.03.020

Leonelli, S. (2016). *Data-Centric Biology: A Philosophical Study*. University of Chicago Press.

Leonelli, S. (2018). The Time of Data: Timescales of Data Use in the Life Sciences. *Philosophy of Science*, *85*(5), 741–754. https://doi.org/10.1086/699699

Leonelli, S. (2019). Data Governance is Key to Interpretation: Reconceptualizing Data in Data Science. *Harvard Data Science Review*, *1*(1). https://doi.org/10.1162/99608f92.17405bb6

Leonelli, S., & Ankeny, R. A. (2012). Re-thinking organisms: The impact of databases on model organism biology. *Studies in History and Philosophy of Science Part C: Studies in History and Philosophy of Biological and Biomedical Sciences*, *43*(1), 29–36. https://doi.org/10.1016/j.shpsc.2011.10.003

Leonelli, S., & Tempini, N. (Eds.). (2020). *Data Journeys in the Sciences*. Springer. https://doi.org/10.1007/978-3-030-37177-7

Lewin, K. (1951). *Field Theory in Social Science: Selected Theoretical Papers*. Harper.

Li, L., & Ludäscher, B. (2023). On the Reusability of Data Cleaning Workflows. *International Journal of Digital Curation*, *17*(1), Article 1. https://doi.org/10.2218/ijdc.v17i1.828

Lin, D., Crabtree, J., Dillo, I., Downs, R. R., Edmunds, R., Giaretta, D., De Giusti, M., L'Hours, H., Hugo, W., Jenkyns, R., Khodiyar, V., Martone, M. E., Mokrane, M., Navale, V., Petters, J., Sierman, B., Sokolova, D. V., Stockhause, M., & Westbrook, J. (2020). The TRUST Principles for digital repositories. *Scientific Data*, *7*(1), Article 1. https://doi.org/10.1038/s41597-020-0486-7

Loukissas, Y. A. (2019). *All Data Are Local: Thinking Critically in a Data-Driven Society*. MIT Press.

Lynch, C. A. (2001). When documents deceive: Trust and provenance as new factors for information retrieval in a tangled web. *Journal of the American Society for Information Science & Technology*, *52*(12), 12–17. http://onlinelibrary.wiley.com/doi/10.1002/1532-2890%282000%2952:1%3C12::AID-ASI1062%3E3.0.CO;2-V/abstract

Lynch, C. A., & Garcia-Molina, H. (1995). *Interoperability, scaling, and the digital libraries research agenda*. IITA Digital Libraries Workshop. http://www-diglib.stanford.edu/diglib/pub/reports/iita-dlw/main.html

Mager, A., & Katzenbach, C. (2021). Future imaginaries in the making and governing of digital technology: Multiple, contested, commodified. *New Media & Society*, *23*(2), 223–236. https://doi.org/10.1177/1461444820929321

*Maintainers*. (2024). The Maintainers, A Global Research and Practice Network. https://themaintainers.org/

Mandell, R. A. (2012). *Researchers' Attitudes towards Data Discovery: Implications for a UCLA Data Registry* [MLIS, University of California, Los Angeles]. http://escholarship.org/uc/item/5bv8j7g3

Marsolek, W., Wright, S. J., Luong, H., Braxton, S. M., Carlson, J., & Lafferty-Hess, S. (2023a). Understanding the value of curation: A survey of researcher perspectives of data curation services from six US institutions. *PLOS ONE*, *18*(11), e0293534. https://doi.org/10.1371/journal.pone.0293534

Marsolek, W., Wright, S. J., Luong, H., Braxton, S. M., Carlson, J., & Lafferty-Hess, S. (2023b). Understanding the value of curation: A survey of researcher perspectives of data curation services from six US institutions. *PLOS ONE*, *18*(11), e0293534. https://doi.org/10.1371/journal.pone.0293534

Martone, M. E., Nakamura, R., & Borgman, C. L. (Eds.). (2022). *Changing the Culture on Data Management and Data Sharing in Biomedicine. Special Theme Issue of Harvard Data Science Review: Vol. 4:3* (Issue 3). Harvard Data Science Review. https://hdsr.mitpress.mit.edu/volume4issue3

Mayernik, M. S. (2016). Research data and metadata curation as institutional issues. *Journal of the Association for Information Science and Technology*, *67*(4), 973–993. https://doi.org/10.1002/asi.23425

Mayernik, M. S. (2019). Metadata accounts: Achieving data and evidence in scientific research. *Social Studies of Science*, *49*(5), 732–757. https://doi.org/10.1177/0306312719863494

Mayernik, M. S. (2020a). Metadata. In B. Hjørland & C. Gnoli (Eds.), *Encyclopedia of Knowledge Organization*. https://www.isko.org/cyclo/metadata

Mayernik, M. S. (2020b). Metadata. *Knowledge Organization*, *47*(8), 696–713. https://doi.org/10.5771/0943-7444-2020-8-696

Mayernik, M. S., & Acker, A. (2018). Tracing the traces: The critical role of metadata within networked communications. *Journal of the Association for Information Science and Technology*, *69*(1), 177–180. https://doi.org/10.1002/asi.23927

Mayernik, M. S., & Liapich, Y. (2022). The Role of Metadata and Vocabulary Standards in Enabling Scientific Data Interoperability: A Study of Earth System Science Data Facilities. *Journal of eScience Librarianship*, *11*(2), Article 2. https://doi.org/10.7191/jeslib.619

Mayernik, M. S., Wallis, J. C., Pepe, A., & Borgman, C. L. (2008, February 28). Whose data do you trust? Integrity issues in the preservation of scientific data. *Proceedings of iConference 2008: iFutures: Systems, Selves, Society*. iConference, Los Angeles, CA. IDEALS. http://hdl.handle.net/2142/15119

McKiernan, E. C., Barba, L., Bourne, P. E., Carter, C., Chandler, Z., Choudhury, S., Jacobs, S., Katz, D. S., Lieggi, S., Plale, B., & Tananbaum, G. (2023). Policy recommendations to ensure that research software is openly accessible and reusable. *PLOS Biology*, *21*(7), e3002204. https://doi.org/10.1371/journal.pbio.3002204

McPhillips, T., Bowers, S., Zinn, D., & Ludäscher, B. (2009). Scientific workflow design for mere mortals. *Future Generation Computer Systems*, *25*(5), 541–551. https://doi.org/10.1016/j.future.2008.06.013

Merrill, D. (2014). *Economic Perspectives for Long-term Digital Preservation*. Hitachi Data Systems.

Meyns, C. (2019). 'Data' in the Royal Society's Philosophical Transactions, 1665–1886. *Notes and Records: The Royal Society Journal of the History of Science*, *74*(3), 507–528. https://doi.org/10.1098/rsnr.2019.0024

Min, B., Ross, H., Sulem, E., Veyseh, A. P. B., Nguyen, T. H., Sainz, O., Agirre, E., Heintz, I., & Roth, D. (2024). Recent Advances in Natural Language Processing via Large Pre-trained Language Models: A Survey. *ACM Computing Surveys*, *56*(2), 1–40. https://doi.org/10.1145/3605943

Moreau, L., & Groth, P. (2013). *PROV-Overview* [W3C Note]. W3C.

Nagaraj, A. (2024). A Mapping Lens for Estimating Data Value. *Harvard Data Science Review*, *Special Issue 4*. https://doi.org/10.1162/99608f92.82f0de5a

NASA/ADS. (2024). *Astrophysics Data System*. https://ui.adsabs.harvard.edu/

National Academies of Sciences. (2019). *Reproducibility and Replicability in Science*. https://doi.org/10.17226/25303

National Academies of Sciences. (2021). *Developing a Toolkit for Fostering Open Science Practices: Proceedings of a Workshop*. The National Academies Press. https://doi.org/10.17226/26308

National Academies of Sciences, Engineering, and Medicine. (2020). *Life-Cycle Decisions for Biomedical Data: The Challenge of Forecasting Costs* (p. 25639). National Academies Press. https://doi.org/10.17226/25639

National Health and Medical Research Council. (2018). *Australian Code for the Responsible Conduct of Research, 2018*. https://www.nhmrc.gov.au/guidelines-publications/r41

*National Information Standards Organization*. (2024). https://www.niso.org/

National Institutes of Health. (2015, April 21). *Support Interoperability of NIH Funded Biomedical Data Repositories*. http://grants.nih.gov/grants/guide/pa-files/PA-15-144.html

National Institutes of Health. (2017). *NIH Sharing Policies and Related Guidance on NIH-Funded Research Resources* [Government - Funding agency]. National Institutes of Health. https://grants.nih.gov/policy/sharing.htm

National Science Board. (2005). *Long-Lived Digital Data Collections*. http://www.nsf.gov/pubs/2005/nsb0540/

Nature Editorial Board. (2021). The next 20 years of human genomics must be more equitable and more open. *Nature*, *590*(7845), 183–184. https://doi.org/10.1038/d41586-021-00328-0

Ndegwa, H., & Kuchma, I. (2024). Repositories transforming scholarly communication: An Insights special collection. *Insights*, *37*(1). https://doi.org/10.1629/uksg.665

Nosek, B. A., & Lakens, D. (2014). Registered Reports: A Method to Increase the Credibility of Published Results. *Social Psychology*, *45*(3), 137–141. https://doi.org/10.1027/1864-9335/a000192

Nyberg Åkerström, W., Baumann, K., Corcho, O., David, R., Le Franc, Y., Madon, B., Magagna, B., Micsik, A., Molinaro, M., Ojsteršek, M., Peroni, S., Scharnhorst, A., Vogt, L., & Widmann, H. (2024). *Developing and implementing the semantic interoperability recommendations of the EOSC Interoperability Framework*. EOSC Association AISBL. https://urn.kb.se/resolve?urn=urn:nbn:se:uu:diva-529669

Office of The Director, National Institutes of Health. (2020). *Final NIH Policy for Data Management and Sharing* (NOT-OD-21-013:). National Institutes of Health. https://grants.nih.gov/grants/guide/notice-files/NOT-OD-21-013.html

O'Grady, C. (2023). Preregistering, transparency, and large samples boost psychology studies' replication rate to nearly 90%. *Science*. https://www.science.org/content/article/preregistering-transparency-and-large-samples-boost-psychology-studies-replication-rate

O'Hara, K., Hall, W., & Cerf, V. (2021). *Four Internets: Data, Geopolitics, and the Governance of Cyberspace*. Oxford University Press.

Olson, G. M., & Olson, J. S. (2000). Distance Matters. *Human–Computer Interaction*, *15*(2–3), 139–178. https://doi.org/10.1207/S15327051HCI1523_4

Olson, G. M., Zimmerman, A. S., & Bos, N. M. (Eds.). (2008). *Scientific Collaboration on the Internet*. MIT Press.

Open Science Foundation. (2024). Addressing Fear of Scooping on OSF Preregistrations. *OSF Monthly Tips and Tricks*. https://mailchi.mp/osf/osf-tips-mar-1388047?e=20b876654d

Ott, F., Elger, K., Frenzel, S., Brauser, A., & Lorenz, M. (2024). *The role of research data repositories for large integrative research infrastructures*. EGU General Assembly, Vienna. https://doi.org/10.5194/egusphere-egu24-19436

*Our World in Data*. (2024). Our World in Data. https://ourworldindata.org

Parmiggiani, E., Amagyei, N. K., & Kollerud, S. K. S. (2023). Data curation as anticipatory generification in data infrastructure. *European Journal of Information Systems*, *0*(0), 1–20. https://doi.org/10.1080/0960085X.2023.2232333

Pasquetto, I. V. (2018). *From Open Data to Knowledge Production: Biomedical Data Sharing and Unpredictable Data Reuses* [Ph.D. Dissertation, UCLA]. https://escholarship.org/uc/item/1sx7v77r

Pasquetto, I. V., Borgman, C. L., & Wofford, M. F. (2019). Uses and Reuses of Scientific Data: The Data Creators' Advantage. *Harvard Data Science Review*, *1*(2). https://doi.org/10.1162/99608f92.fc14bf2d

Pasquetto, I. V., Randles, B. M., & Borgman, C. L. (2017). On the Reuse of Scientific Data. *Data Science Journal*, *16*. https://doi.org/10.5334/dsj-2017-008

Pasquier, T., Lau, M. K., Trisovic, A., Boose, E. R., Couturier, B., Crosas, M., Ellison, A. M., Gibson, V., Jones, C. R., & Seltzer, M. (2017). If these data could talk. *Scientific Data*, *4*, 170114. https://doi.org/10.1038/sdata.2017.114

Publications Office of the European Union. (2021). *Data.europa.eu data quality guidelines.* Publications Office. https://data.europa.eu/doi/10.2830/333095

Puschmann, C., & Burgess, J. (2014). Big Data, Big Questions| Metaphors of Big Data. *International Journal of Communication*, *8*, Article 0. https://ijoc.org/index.php/ijoc/article/view/2169

Ribes, D., & Bowker, G. C. (2008). Organizing for Multidisciplinary Collaboration: The Case of the Geosciences Network. In G. M. Olson, A. S. Zimmerman, & N. Bos (Eds.), *Scientific Collaboration on the Internet* (pp. 311–330). MIT Press.

Ribes, D., Hoffman, A. S., Slota, S. C., & Bowker, G. C. (2019). The logic of domains. *Social Studies of Science*. https://doi.org/10.1177/0306312719849709

Roberts, A. S., Alec Glassford, Ash Ngu, Brandon. (2013, May 9). *Commoncrawl Foundation—Nonprofit Explorer*. ProPublica. https://projects.propublica.org/nonprofits/organizations/261635908

Rocca-Serra, P., Gu, W., Ioannidis, V., Abbassi-Daloii, T., Capella-Gutierrez, S., Chandramouliswaran, I., Splendiani, A., Burdett, T., Giessmann, R. T., Henderson, D., Batista, D., Emam, I., Gadiya, Y., Giovanni, L., Willighagen, E., Evelo, C., Gray, A. J. G., Gribbon, P., Juty, N., … Sansone, S.-A. (2023). The FAIR Cookbook—The essential resource for and by FAIR doers. *Scientific Data*, *10*(1), Article 1. https://doi.org/10.1038/s41597-023-02166-3

Rogers, E. M. (1995). *Diffusion of Innovations* (4th ed.). The Free Press.

Rosenberg, D. (2013). Data before the Fact. In L. Gitelman (Ed.), *"Raw Data" is an Oxymoron* (pp. 15–40). MIT Press.

Rosenberg, D. (2018). Data as Word. *Historical Studies in the Natural Sciences*, *48*(5), 557–567. https://doi.org/10.1525/hsns.2018.48.5.557

Russell, A. L., & Vinsel, L. (2018). After Innovation, Turn to Maintenance. *Technology and Culture*, *59*(1), 1–25. https://doi.org/10.1353/tech.2018.0004

Ryle, G. (1949). *The Concept of Mind*. Hutchinson.

Sansone, S.-A., & Rocca-Serra, P. (2016). *Review: Interoperability standards*. https://doi.org/10.6084/m9.figshare.4055496.v1

Schmidt, K. (2012). The Trouble with 'Tacit Knowledge.' *Computer Supported Cooperative Work (CSCW)*, *21*(2–3), 163–225. https://doi.org/10.1007/s10606-012-9160-8

Serghiou, S., Axfors, C., & Ioannidis, J. P. A. (2023). Lessons learnt from registration of biomedical research. *Nature Human Behaviour*, 1–4. https://doi.org/10.1038/s41562-022-01499-0

Servilla, M., Brunt, J., Costa, D., McGann, J., & Waide, R. (2016). The contribution and reuse of LTER data in the Provenance Aware Synthesis Tracking Architecture (PASTA) data repository. *Ecological Informatics*, *36*, 247–258. https://doi.org/10.1016/j.ecoinf.2016.07.003

Sofi, I. A., Bhat, A., & Gulzar, R. (2024). Global status of dataset repositories at a glance: Study based on OpenDOAR. *Digital Library Perspectives*, *40*(2), 330–347. https://doi.org/10.1108/DLP-11-2023-0094

Star, S. L. (1999). The Ethnography of Infrastructure. *American Behavioral Scientist*, *43*(3), 377–391. https://doi.org/10.1177/00027649921955326

Star, S. L., & Griesemer, J. R. (1989). Institutional Ecology, `Translations' and Boundary Objects: Amateurs and Professionals in Berkeley's Museum of Vertebrate Zoology, 1907-39. *Social Studies of Science*, *19*(3), 387–420. https://doi.org/10.1177/030631289019003001

Star, S. L., & Ruhleder, K. (1996). Steps Toward an Ecology of Infrastructure: Design and Access for Large Information Spaces. *Information Systems Research*, *7*(1), 111–134. https://doi.org/10.1287/isre.7.1.111

Staunton, C., Barragán, C. A., Canali, S., Ho, C., Leonelli, S., Mayernik, M., Prainsack, B., & Wonkham, A. (2021). Open science, data sharing and solidarity: Who benefits? *History and Philosophy of the Life Sciences*, *43*(4), 115. https://doi.org/10.1007/s40656-021-00468-6

*The Dataverse Project*. (2024, July 3). https://dataverse.org/home

The Information Maintainers, Olson, D., Meyerson, J., Parsons, M. A., Castro, J., Lassere, M., Wright, D. J., Arnold, H., Galvan, A. S., Hswe, P., Nowviskie, B., Russell, A., Vinsel, L., & Acker, A. (2019). *Information Maintenance as a Practice of Care*. https://doi.org/10.5281/zenodo.3236410

Thomer, A. K., Wickett, K. M., Baker, K. S., Fouke, B. W., & Palmer, C. L. (2018). Documenting provenance in noncomputational workflows: Research process models based on geobiology fieldwork in Yellowstone National Park. *Journal of the Association for Information Science and Technology*, *69*(10), 1234–1245. https://doi.org/10.1002/asi.24039

Treloar, A., & Klump, J. (2019). Updating the Data Curation Continuum: Not just Data, still focussed on Curation, more about Domains. *International Journal of Digital Curation*, *14*(1), 87–101. https://doi.org/10.2218/ijdc.v14i1.643

Uffen, H., & Kinkel, T. (2019). *Controlling the cost of long-term digital accessibility. A cost model for long-term digital accessibility*. https://doi.org/10.5281/ZENODO.4274258

*UK Data Archive*. (2024). https://www.data-archive.ac.uk/

Van de Sompel, H., Lagoze, C., Bekaert, J., Liu, X., Payette, S., & Warner, S. (2006). An interoperable fabric for scholarly value chains. *D-Lib Magazine*, *12*. http://www.dlib.org/dlib/october06/vandesompel/10vandesompel.html

Van de Sompel, H., & Nelson, M. L. (2015). Reminiscing About 15 Years of Interoperability Efforts. *D-Lib Magazine*, *21*(11/12). http://dx.doi.org/10.1045/november2015-vandesompel

Veitch, D. P., Weiner, M. W., Aisen, P. S., Beckett, L. A., DeCarli, C., Green, R. C., Harvey, D., Jack Jr., C. R., Jagust, W., Landau, S. M., Morris, J. C., Okonkwo, O., Perrin, R. J., Petersen, R. C., Rivera-Mindt, M., Saykin, A. J., Shaw, L. M., Toga, A. W., Tosun, D., … Initiative, A. D. N. (2022). Using the Alzheimer's Disease Neuroimaging Initiative to improve early detection, diagnosis, and treatment of Alzheimer's disease. *Alzheimer's & Dementia*, *18*(4), 824–857. https://doi.org/10.1002/alz.12422

Verburg, M., Dillo, I., & Grootveld, M. (2024). *Uniting researchers and repositories to improve data FAIRness: The Netherlands in a European perspective*. https://doi.org/10.5281/zenodo.10534087

Vertesi, J., & Dourish, P. (2011). The value of data: Considering the context of production in data economies. *Proceedings of the ACM 2011 Conference on Computer Supported Cooperative Work*, 533–542. https://doi.org/10.1145/1958824.1958906

Wallis, J. C., Mayernik, M. S., Borgman, C. L., & Pepe, A. (2010). Digital libraries for scientific data discovery and reuse: From vision to practical reality. *Proceedings of the 10th Annual Joint Conference on Digital Libraries*, 333–340. https://doi.org/10.1145/1816123.1816173

Wallis, J. C., Rolando, E., & Borgman, C. L. (2013). If we share data, will anyone use them? Data Sharing and reuse in the long tail of science and technology. *PLoS ONE*, *8*. https://doi.org/10.1371/journal.pone.0067332

Wang, X., Duan, Q., & Liang, M. (2021). Understanding the process of data reuse: An extensive review. *Journal of the Association for Information Science and Technology*, *72*(9), 1161–1182. https://doi.org/10.1002/asi.24483

Weber, N. M., Palmer, C. L., & Chao, T. C. (2012). Current Trends and Future Directions in Data Curation Research and Education. *Journal of Web Librarianship*, *6*(4), 305–320. https://doi.org/10.1080/19322909.2012.730358

Weeden, K. A. (2023). Crisis? What Crisis? Sociology's Slow Progress Toward Scientific Transparency. *Harvard Data Science Review*, *5*(4). https://doi.org/10.1162/99608f92.151c41e3

Weller, S. (2023). Fostering habits of care: Reframing qualitative data sharing policies and practices. *Qualitative Research*, *23*(4), 1022–1041. https://doi.org/10.1177/14687941211061054

Wilhelm, E. E., Oster, E., & Shoulson, I. (2014). Approaches and Costs for Sharing Clinical Research Data. *JAMA*, *311*(12), 1201. https://doi.org/10.1001/jama.2014.850

Wilkinson, M. D., Dumontier, M., Aalbersberg, Ij. J., Appleton, G., Axton, M., Baak, A., Blomberg, N., Boiten, J.-W., da Silva Santos, L. B., Bourne, P. E., Bouwman, J., Brookes, A. J., Clark, T., Crosas, M., Dillo, I., Dumon, O., Edmunds, S., Evelo, C. T., Finkers, R., … Mons, B. (2016). The FAIR Guiding Principles for scientific data management and stewardship. *Scientific Data*, *3*, 160018. https://doi.org/10.1038/sdata.2016.18

Wing, J. M. (2019). The Data Life Cycle. *Harvard Data Science Review*, *1*(1). https://doi.org/10.1162/99608f92.e26845b4

Wyatt, S. (2021). Metaphors in critical Internet and digital media studies. *New Media & Society*, *23*(2), 406–416. https://doi.org/10.1177/1461444820929324

Wylie, C. D. (2024). Timing science: The temporal role of scientists in the construction of data. *Philosophy, Theory, and Practice in Biology*, *16*(2), Article 2. https://doi.org/10.3998/ptpbio.5646

Yehudi, Y., Hughes-Noehrer, L., Goble, C., & Jay, C. (2023). Subjective data models in bioinformatics and how wet lab and computational biologists conceptualise data. *Scientific Data*, *10*(1), Article 1. https://doi.org/10.1038/s41597-023-02627-9

*Zenodo*. (2024). https://zenodo.org/

Zhang, H., Santos, A., & Freire, J. (2021). DSDD: Domain-Specific Dataset Discovery on the Web. *Proceedings of the 30th ACM International Conference on Information & Knowledge Management*, 2527–2536. https://doi.org/10.1145/3459637.3482427

Zhang, P., Gregory, K. M., Yoon, A., & Palmer, C. (2023). Conceptualizing Data Behavior: Bridging Data-centric and User-centric Approaches. *Proceedings of the Association for Information Science and Technology*, *60*(1), 856–860. https://doi.org/10.1002/pra2.878

Zhang, Q. P., Olson, G. M., & Olson, J. S. (2004). Does video matter more for long distance collaborators? *International Journal of Psychology*, *39*, 360–360.

Zimmerman, A. S. (2007). Not by metadata alone: The use of diverse forms of knowledge to locate data for reuse. *International Journal on Digital Libraries*, *7*(1–2), 5–16. https://doi.org/10.1007/s00799-007-0015-8

Zooniverse. (2024). *Old Weather Collections*. https://www.zooniverse.org/projects/zooniverse/old-weather/collections